\documentclass{nature}
\usepackage{soul}
\usepackage{xcolor}%

\usepackage{lineno}
\usepackage{graphicx}%
\usepackage{capt-of}%

\usepackage{amsmath}
\usepackage{amssymb}
\title{Spin-polarized Correlated Insulator and Superconductor in Twisted Double Bilayer Graphene}

\author{Xiaomeng Liu$^{1\dagger*}$, Zeyu Hao$^{1\dagger}$, Eslam Khalaf$^{1}$, Jong Yeon Lee$^{1}$, Kenji Watanabe$^2$, Takashi Taniguchi$^2$, Ashvin Vishwanath$^{1}$ \& Philip Kim$^{1*}$}

\begin{document}

\maketitle

\begin{affiliations}
 \item Department of Physics, Harvard University, Cambridge, Massachusetts 02138, USA
 \item National Institute for Material Science, 1-1 Namiki, Tsukuba 305-0044, Japan
 \\$\dagger$Theses authors contribute equally to this work.
\end{affiliations}

\begin{abstract}
Ferromagnetism and superconductivity typically compete with each other~\cite{Khan1985} since the internal magnetic field generated in a magnet suppresses the formation of spin-singlet Cooper pairs in conventional superconductors~\cite{tinkham_introduction_2004}. Only a handful of  ferromagnetic superconductors are known in heavy fermion systems~\cite{saxena_superconductivity_2000, aoki_coexistence_2001, huy_superconductivity_2007}, where many-body electron interactions promoted by the narrow energy bands play a key role in stabilizing  these emergent states. Recently, interaction-driven superconductivity and ferromagnetism have been demonstrated as separate phenomena in different density regimes of flat bands enabled by graphene moir\'e superlattices~\cite{cao_correlated_2018, cao_unconventional_2018, yankowitz_tuning_2019, sharpe_emergent_2019}. Combining superconductivity and magnetism in a single ground state may lead to more exotic quantum phases. Here, employing van der Waals heterostructures of twisted double bilayer graphene (TDBG), we realize a flat electron band that is tunable by perpendicular electric fields. Similar to the magic angle twisted bilayer graphene, TDBG exhibits energy gaps at the half and quarter filled flat bands, indicating the emergence of correlated insulating states. We find that the gaps of these insulating states increase with in-plane magnetic field, suggesting a ferromagnetic order. Upon doping the ferromagnetic half-filled insulator, superconductivity emerges with a critical temperature controlled by both density and electric fields. We observe that the in-plane magnetic field enhances the superconductivity in the low field regime, suggesting spin-polarized electron pairing. Spin-polarized superconducting states discovered in TDBG provide a new route to engineering interaction-driven topological superconductivity~\cite{Bernevig}.
\end{abstract}

Moir\'e superlattices of two dimensional (2D) van der Waals (vdW) materials provide a new scheme for creating correlated electronic states. By stacking atomically thin vdW layers with a controlled twist angle $\theta$ between them, the size of the moir\'e unit cell can be tuned~\cite{dean_hofstadters_2013,ponomarenko_cloning_2013, hunt_massive_2013}. In particular, in twisted bilayer graphene (TBG), the weak interlayer coupling can open up energy gaps at the boundary of the mini-Brillouin zone, which modifies the energy band-width of the coupled system.  Theoretically, it has been predicted that around $\theta \approx$1.1$^{\circ}$ (the so-called magic-angle (MA)), the interlayer hybridization induces isolated flat bands with drastically reduced bandwidth and enhanced density of states~\cite{bistritzer_moire_2011}. Recent experiments performed in MA-TBG indeed confirmed the appearance of correlated insulating states associated with the flat band~\cite{cao_correlated_2018}. Surprisingly, it has also been discovered that upon doping the half-filled insulator in MA-TBG, superconductivity appears, phenomenologically resembling high temperature cuprate superconductors whose undoped parent compounds are Mott insulators~\cite{keimer_quantum_2015}. 

In TBG, the value of the MA is governed by the individual layer band structure and interlayer coupling strength. While the mechanism of superconductivity in the moir\'e superlattice is under debate~\cite{xu_topological_2018, po_origin_2018, koshino_maximally_2018, kang_symmetry_2018,lian_twisted_2018, Fu}, continuous tuning of the superlattice band structure across the MA condition will provide experimental control to elucidate the correlated electronic states. However, to date, such an experimental control is largely achieved by fabricating samples with different twist angles, and only very limited tunability has been demonstrated in TBG by application of hydrostatic pressure ~\cite{yankowitz_tuning_2019, carr_pressure_2018}. In ABC trilayer graphene/hBN superlattice, electric field was shown to modulate the correlated insulator gap~\cite{chen_evidence_2019}, opening up the possibility of easy tuning of the moir\'e flat band with electric field. Here, we demonstrate electric field-tunable moir\'e flat bands in twisted double bilayer graphene (TDBG), consisting of two Bernal-stacked bilayer graphene sheets misaligned with a twist angle (Fig. 1a). Unlike monolayer graphene, the band structure of Bernal-stacked bilayer graphene itself can be tuned by a perpendicular displacement field $D$~\cite{castro_neto_electronic_2009}. As $|D|$ increases, the parabolic band touching at charge neutrality of bilayer graphene opens up a gap and the bottom (top) of the conduction (valance) band develops a shallow Mexican-hat energy dispersion distorted by trigonal warping~\cite{mccann_electronic_2013}. The gap in the bilayer graphene can be as big as 200~meV for large $|D|$ before the gate dielectric breaks down~\cite{zhang_direct_2009}. In TDBG, where two bilayers are stacked, this additional band adjustment changes the energy dispersion of each constituent bilayer, allowing a new experimental knob to tune the flat band condition. Fig.~1c shows rearranged moir\'e band structures computed at finite $D$ using the continuum model approximation~\cite{bistritzer_moire_2011, zhang_nearly_2019 ,chebrolu_flatbands_2019,choi_intrinsic_2019}. We find that well-isolated narrow conduction band can appear in a range of twist angle $\theta$ 
%$1.5$^{\circ}$f
, where the inter-band energy gaps and bandwidth can be controlled by the displacement field (see Supplementary Information (SI) S1 for detail). 

We fabricated TDBG devices by tearing and stacking Bernal stacked bilayer graphene~\cite{cao_superlattice-induced_2016, kim_van_2016}. Separated by hexagonal boron nitride from TDBG, top and bottom graphite gates with voltages $V_{TG}$ and $V_{BG}$ are used to control the density of electrons $n$ and $D$ independently: $n=C_{TG}V_{TG}+C_{BG}V_{BG}, D=(C_{TG}V_{TG}-C_{BG}V_{BG})/2$, where $C_{TG}$ ($C_{BG}$) is the capacitance between TDBG and the top (bottom) gate. Fig. 1d shows the four-probe resistivity $\rho$ measured in the TDBG with $\theta=1.33^\circ$ as a function of $V_{TG}$ and $V_{BG}$ at temperature $T=1.6$~K. Generally, $\rho$ exhibits several insulating states where the corresponding conductance $\sigma=\rho^{-1}$ vanishes as temperature $T$ decreases (Fig.~1g), suggesting gap opening at the Fermi level of the system. Some insulating regions identified in Fig.~1d can be explained in the single particle band structure described in Fig.~1c. For example, we find that the charge neutral point (CNP) is gapless at $D=0$ but a gap develops for $|D|>D_1 \neq 0$. Similarly, at full moir\'e band filling $n=\pm n_s$, where $n_s$ is the density corresponding to four electrons per moir\'e unit cell considering spin and valley degeneracy, energy gaps $\Delta_{\pm n_s}$ are present within displacement field ranges $|D|<D^{\pm}_2$. The ranges of $D^{\pm}_2$ are different in the conduction band ($+$) and valance ($-$) band, due to the lack of electron-hole symmetry of TDBG. We further confirm the narrowness of the band by estimating the effective cyclotron mass $m^*$ from the temperature dependent magnetoresistance oscillation measurement (see SI, S3.3). Fig.~1f shows that $m^*\approx 0.2 m_e$ for the first valance band (v$_1$) and  $\sim 0.4 m_e$ for the first conduction band (c$_1$), where $m_e$ is the bare electron mass. Considering the effective mass of Bernal-stacked bilayer graphene is $\approx 0.04m_e$ ~\cite{li_effective_2016}, the experimentally observed large $m^*$ indicates an order of magnitude narrower band width than folded bilayer graphene bands especially for c$_1$ band in this regime of gate configurations. Calculations based on a continuum model (Fig. 1c) demonstrate the existence of a isolated flat band at this twist angle under displacement field (theoretical bandwidth is around 10$\sim$15~meV), providing a qualitative agreement with the experimental results (SI, S1). 

In the single particle band structure, we expect a narrow but uninterrupted spectrum within the lowest moir\'e conduction band (c$_1$), separated by band gaps from both the valance band (v$_1$) and higher conduction band (c$_2$) for $D_1<|D|<D^+_2$. However, we observe development of well-defined insulating behavior at half-filling $n=n_s/2$ (Fig. 1d, f). The onset of this insulating state coincides with $D_1$. However, this state ends well before $D$ reaches $D^+_2$, suggesting both the isolation and flatness of the band are responsible for creating the observed correlated gap (SI, S2). The absence of correlated insulating behavior in the hole-doped regime in similar experimental conditions can also be explained by the wider band width in the moir\'e valance band v$_2$ than that of c$_1$ (SI, S1). 

We measure the size of the gaps from temperature-dependent activating behavior of $\rho$ (Fig.~1g inset). For $\theta=1.33^\circ$ TDBG, the half-filled insulator is robust with an energy gap of $\Delta_{n_s/2}=3$~meV and persists up to a perpendicular magnetic field $B_\perp \approx 7$~T (SI, S3.2). Since c$_1$ band is spin and valley degenerate in a single particle picture, the correlated insulator we observed in the half-filling is likely polarized in the spin-valley space. Applying in-plane magnetic field $B_\parallel$ can probe the spin structure of the state without significant coupling to the valley. In MA-TBG, it has been shown that $B_\parallel$ reduces $\Delta_{n_s/2}$, suggesting that the half-filled insulator is a spin-unpolarized state. Fig.~1e shows the change of $\rho$ as a function of $B_\parallel$ in our TDBG sample. We find that the half-filled insulating state becomes more insulating as $B_\parallel$ increases (Fig.~1g inset) and the displacement field range where the half-filled insulator spans also enlarges (Fig.~1e). More quantitatively, we find the activation gap $\Delta_{n_s/2}$ increases with $B_\parallel$ for the half-filled insulating state. The growth of $\Delta_{n_s/2}$ roughly follows the Zeeman energy scale $g\mu_B B_\parallel$ where $\mu_B$ is the Bohr magneton and the effective g-factor $g=2$ (dotted line in Fig.~1h). This observation is consistent with a picture where the occupied states (half of the states in c$_1$) are spin-polarized along the direction of the external magnetic field. The unoccupied states then carry the opposite spin, separated by a spontaneous ferromagnetic gap. Considering spin-1/2, the Zeeman term lowers the energy of filled states $\Delta E_{\downarrow}= -g \mu_{B} B/2$, while boosting the energy of empty states with opposite spins by $\Delta E_{\uparrow}= g \mu_{B} B/2$, pushing the two bands further apart and enhancing the gap (as illustrated by Fig.~1h insets).

Applying $B_\parallel$ also induces additional correlated insulating states at quarter-filling ($n=\frac{1}{4}n_s$) and three-quarter-filling ($n=\frac{3}{4}n_s$), which are absent at zero magnetic field. As shown in the upper panel of Fig.~1e, these insulating states are more fragile than the half-filled insulating state as they are stabilized in narrower displacement field ranges. The quarter-filled insulating gap opens at $B_\parallel \approx 4$~T and increases as $B_\parallel$ increases (Fig.~1h). Considering that the full bands are spin and valley degenerate, these quarter-filled (or quarter-empty) states are presumably polarized in spin-valley phase space. While the details of the mechanism and the texture of the spin/ valley structure is yet to be investigated, the fact that the energy gaps corresponding to these states increase approximately with $g\mu_B B_\parallel$ suggests that the corresponding insulating states are spin polarized as illustrated in the upper inset of Fig.~1h.

Fig.~2a shows $\rho$ measured in another sample with smaller twist angle $\theta=1.24^\circ$. Here, again, the half-filled insulating regime is found at finite displacement fields. However, in this smaller angle, we find the correlated insulating behavior is weaker with a much smaller energy gap ($\Delta_{n_s/2}=0.3$~meV) than that of the $\theta=1.33^\circ$ sample. Despite an order of magnitude reduction compared to the previous sample,  $\Delta_{n_s/2}$  increases with increasing $B_\parallel$ closely following the Zeeman energy, $g \nu_B B_\parallel$, indicating spin-polarization of the c$_1$ band (Fig.~3b). We also note that the single particle gap $\Delta_{n_S}$ between c$_1$ and c$_2$ decreases linearly with Zeeman energy with g-factor of 2 (Fig.~3b). This behavior can be explained by splitting and down-shifting of spin-aligned states in $c_2$ and up-shifting of spin-anti-aligned states in c$_1$ , providing additional confirmation of spin-polarization of the c$_1$ band, as illustrated in Fig.~3b inset. 

Upon doping electrons in the half-filled insulator, we discover superconductivity in the $\theta=1.24^\circ$ sample. Fig.~2b shows the temperature dependent $\rho$ measured at several different gate configurations marked in Fig.~2a with corresponding colored symbols. We find that the superconductivity appears within the upper right side of the half-filled insulator region in Fig.~2a, surrounded by a half-circle ring of high $\rho$ values ($\gtrsim 0.5$~k$\Omega$). In the center of this superconducting half-circle region ($D=D_m\approx$ 0.38~V/nm and $n=n_m\approx 2n_s/3$), the observed superconductivity is strongest: $\rho(T)$ rapidly drops around $T=6$~K and reaches zero at $T\approx 3.5$~K (black line in Fig.~2b). At finite bias, a typical current-voltage ($I$--$V$) curve for superconductors is observed (Fig.~2c). By fitting the $I$--$V$ curves with $V \propto I^{\alpha}$, we extract the Berezinskii-Kosterlitz-Thouless (BKT) transition temperature $T_{BKT}\approx 3.5$~K, defined at $\alpha(T)=3$ (Fig.~2c inset)~\cite{tinkham_introduction_2004}.

We find that the superconductivity in TDBG sample can be tuned by both density and displacement field. In Fig.~2d, we show $\rho(T)$ as a function of $n$ at a fixed $D=D_m$. A dome region of superconductivity can be recognized next to the half-filled insulator ($n=n_s/2$), similar to the one seen in MA-TBG, but with a much larger temperature scale. At the optimal doping ($n=n_m$), the superconductivity observed in TDBG also weakens away from the optimal displacement field $D_m$. In Fig.~2b inset, $\rho(D,T)$ exhibits a dome-like domain of superconducting region in the $(D, T)$ plane. Inside of this superconducting dome region, we find that the differential resistance $dV/dI$ responds sharply to increasing temperature, magnetic field and drive current, which are signatures of superconductivity. Outside of the superconducting region, however, we do not observe sharp temperature transition nor typical superconducting $I$--$V$ behavior. The low resistance or even negative resistance that appeared at some gate configurations (as shown in Fig. 3a) is possibly due to the mesoscopic ballistic transport (see SI S5.2). 

On applying perpendicular magnetic fields $B_{\perp}$, the superconductivity is suppressed above $B_{\perp}^c=0.1$~T (Fig.~2e).  For $B_{\perp}\gg B_{\perp}^c$, we observe that $\rho(n,B_{\perp})$ exhibits Shubnikov-de Haas (SdH) quantum oscillations at temperatures above the superconducting dome region where the superconductivity was destroyed. The corresponding Landau fan can be traced to the electron filling of the half-filled insulating state. This observation indicates that the TDBG superconductivity is closely linked with the ferromagnetic half-filled insulating states. Thus the Cooper pairs responsible for the superconductivity are likely built upon a Fermi surface that is derived from the doped ferromagnetic insulator. 

In order to explore the spin structure of the TDBG superconductivity, we investigate the behavior of $\rho(T)$ as a function of $B_{\parallel}$. Fig.~3a shows that the superconducting dome in $(n,B_{\parallel})$ plane with the maximum critical parallel magnetic field $B_{\parallel}^c\approx 1$~T. The salient experimental feature is the $B_{\parallel}$ dependence of the superconducting state below the critical field $B_{\parallel}^c$. Fig.~3c shows $\rho$ at the optimal superconducting state $(n_m,D_m)$ measured as a function of $T$ and $B_\parallel$. In this optimal superconducting state, $\rho$ vanishes critically as $T$ and $B_{\parallel}$ decreases. We use a phenomenological definition of the critical temperature with the 50\% transition point defined as $T_{50\%}$. Interestingly, $T_{50\%}(B_{\parallel})$ follows a non-monotonic behavior. In particular,  $T_{50\%}$ increases as $B_{\parallel}$ increases from 0 to $\sim$0.3~T before it decreases for $B_{\parallel}>0.3$~T. We also note that under large in-plane magnetic fields $B_\parallel>1.5$~T, an insulating behavior can be recognized at low temperatures, where $d\rho/dT < 0$ and $\rho >$ 5~k$\Omega$ (see SI S5.4). 

Our  observation of the strengthening of superconductivity with increasing magnetic field is extremely unusual within the framework of spin-singlet s-wave superconductivity. Only a handful of examples of related experimental observations are reported in superconductivity based on heavy fermions ~\cite{saxena_superconductivity_2000, aoki_coexistence_2001, huy_superconductivity_2007} and topological systems with strong spin-orbit coupling~\cite{sajadi_gate-induced_2018}, where complex spin textures of electronic systems are involved. To further quantify the enhanced superconductivity at the low $B_{\parallel}$ regime, we performed $I$--$V$ characterization at the optimal gate configuration ($n_m$, $D_m$) as a function of $B_{\parallel}$ and $T$ to obtain $T_{BKT}$. Similar to $T_{50\%}$ above, $T_{BKT}(B_{\parallel})$ also exhibits a non-monotonic behavior as shown in Fig.~3c (black circles). At low magnetic regime, particularly, we find $k_B \Delta T_{BKT}\approx g\mu_B |B_{\parallel}|$, developing a sharp cusp at $B_{\parallel}=0$. This is in reminiscent of the magnetic field dependence of the transition temperature in superfluid He$_3$ where spin triplet pairing is obtained ~\cite{ambegaokar_thermal_1973}. The underlying reason for this linear increase in $T_{\rm BKT}$ is the increase in the density of states of the favorably aligned spins due to Zeeman splitting. In general, the resulting linear coefficient depends on details such as Fermi energy and magnetic susceptibility ~\cite{ambegaokar_thermal_1973}. The measured value of this coefficient $\approx g\mu_B/k_B$, which is seemingly independent of these details, hints at the existence of a proximate ferromagnetic quantum critical point \cite{TDBGtheory2019}.

The increase of $T_{BKT}$ with $B_{\parallel}$ suggests the Cooper pairs responsible for the TDBG superconductivity are likely to be spin-polarized. Adding the requirement of overall anti-symmetric wavefunction for electron pairs, the orbital component (including valley) of the Cooper pair wavefunction thus has to be anti-symmetric. One possible scenario for such a state is illustrated in Fig. 3d, where the Cooper pairs form between Fermi surfaces with the same spin (spin-triplet) and opposite valley. If the wavefunction is a valley singlet, then the extended spatial wavefunction should be s- or d-wave. Conversely, with valley-triplet pairing, p-wave superconductivity is then expected. In this ferromagnetic superconductor, parallel magnetic field enlarges the majority spin Fermi surface, and strengthens the superconductivity, inducing the change of critical temperature $\Delta T_c \propto B$. The eventual destruction of superconductivity at high magnetic field can result from the following mechanism.  Magnetic flux in between layers leads to a momentum shift which has opposite sign in the two valleys, thereby bringing the two pairing Fermi surfaces out of alignment. The latter effect is expected to reduce the critical temperature, $\Delta T_c\propto -B^2$~\cite{TDBGtheory2019}. Alternatively, if the ferromagnetic pairing is caused by spin fluctuations, as suggested in the heavy fermion metals~\cite{saxena_superconductivity_2000, aoki_coexistence_2001, huy_superconductivity_2007}, strong parallel magnetic field can suppress the superconductivity by  suppressing spin fluctuations~\cite{julian_viewpoint:_2012}.

\begin{methods}
All the devices presented in this study are prepared with the dry transfer method \cite{Wang2013}, using stamps consisting of polypropylene carbonate (PPC) film and polydimethylsiloxane (PDMS). Half of a bilayer graphene flake is teared and picked up by a stack of graphite/hBN on the transfer stamp. Then the remaining bilayer graphene flake on the substrate is rotated by the desired angle and picked up. The stacks are deposited on 300~nm SiO$_2$/Si substrate after picking up the rest of hBN and graphite layers. Part of the bilayer graphene flakes are extended outside the hBN area onto PPC to prevent graphene from freely rotating on hBN. Resulting stacks are fabricated into 1~$\mu$m wide devices to ensure uniform twist angle in the relative narrow channel. The temperature of the stack is always kept below 180$^\circ$C during stacking and fabrication processes. 

The 1.33$^\circ$ device and 2$^\circ$ device are shaped into Hall bar shaped multi-terminal devices. The graphite top and bottom gates are employed to control the carrier density and displacement field in the channel. In these devices, the silicon back gate is utilized to increase the contact transparency by heavily doping the lead region of the sample with the same type of carriers as in the channel region. In the 1.33$^\circ$ device, top hBN dielectric is 48~nm thick and the bottom hBN thickness is 57~nm. The 1.24$^\circ$ device is shaped into Van der Pauw geometry with a graphite top gate and the silicon substrate bottom gate. The top hBN dielectric for this device is 48~nm thick. The resistivity presented here are measured with 0.5-1~mV voltage bias with a current-limiting-resistor of 100~k$\Omega$ connected in series (which limit the current to 5-10 nA) at 17.7Hz using standard lock-in technique. Four-terminal voltage and source-drain current are measured simultaneously with two lock-in amplifiers to obtain four-terminal resistance. Resistivitiy is then obtained from resistance by multiplying with a geometric factor (one for the Hall bar devices, $\sim$4.5 for the Van der Pauw device). 
%%Put methods in here.  If you are going to subsection it, use
%%\verb|\subsection| commands.  Methods section should be less than
%%800 words and if it is less than 200 words, it can be incorporated
%%into the main text.
\end{methods}

%% Put the bibliography here, most people will use BiBTeX in
%% which case the environment below should be replaced with
%% the \bibliography{} command.
\subsection{Reference}
\bibliography{naturebib.bib}

%% Here is the endmatter stuff: Supplementary Info, etc.
%% Use \item's to separate, default label is "Acknowledgements"

\begin{addendum}
 \item The major experimental work is supported by DOE (DE-SC0012260). P.K. acknowledges partial support from the Gordon and Betty Moore Foundation's EPiQS Initiative through Grant GBMF4543 and the DoD Vannevar Bush Faculty Fellowship N00014-18-1-2877. A.V, J.Y.L and E.K were supported by a Simons Investigator Fellowship. K.W. and T.T. acknowledge support from the Elemental Strategy Initiative
conducted by the MEXT, Japan, A3 Foresight by JSPS and the CREST
(JPMJCR15F3), JST. A portion of this work was performed at the National High Magnetic Field Laboratory, which is supported by the National Science Foundation Cooperative Agreement No. DMR-1157490* and the State of Florida. Nanofabrication was performed at the Center for Nanoscale Systems at Harvard, supported in part by an NSF NNIN award ECS-00335765. We thank Shiang Fang, Stephen Carr, Yonglong Xie, Efthimios Kaxiras, Bertrand Halperin and Jonah Waissman for helpful discussion.

 \item[Competing Interests] The authors declare that they have no
competing financial interests.

\item[Author Comtributions] X.L. and P.K. conceived the experiment. X.L. and Z.H. fabricated the samples, performed the measurements and analyzed the data. E.K., J.Y.L. and A.V. conducted the theoretical analysis. X.L., Z.H. and P.K. wrote the paper with input from E.K., J.Y.L. and A.V. K.W. and T.T. supplied hBN crystals.

 \item[Correspondence] Correspondence and requests for materials
should be addressed to X. Liu~(email: xiaomeng@princeton.edu) and P. Kim~(email: pkim@physics.harvard.edu).
\end{addendum}

%%
%% TABLES
%%
%% If there are any tables, put them here.
%%

\clearpage
%begin{center}

\begin{figure}
\includegraphics[width=\textwidth]{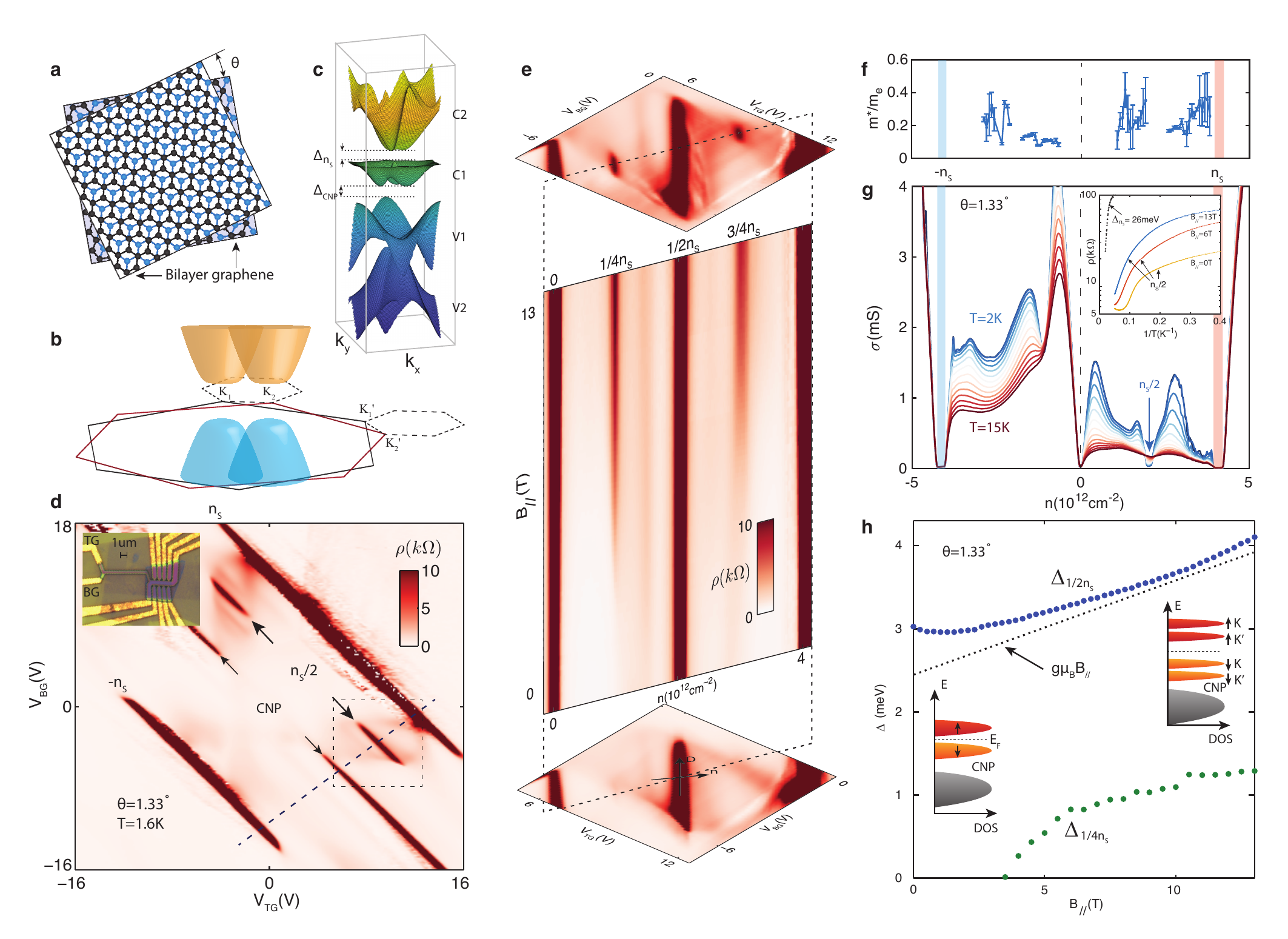}
\end{figure}

\captionof{figure}{\textbf{$\vert$ Half-filled and quarter-filled insulating states in $\theta=1.33^\circ$ sample.} \textbf{a,} Schematic diagram of twisted bilayer-bilayer graphene with twist angle $\theta$. \textbf{b,} Brillouin zone and band structure of the two individual bilayer graphene under perpendicular displacement field. The dashed hexagons represent the mini-Brillouin zone of the moir\'e superlattice. \textbf{c,} Calculated band structure for $\theta=1.33^\circ$ TDBG at optimal displacement field $D_m$. \textbf{d,} Resistivity as a function of top and bottom gate voltages. Charge neutral point (CNP), full-filled band ($n_s$) and half-filled band ($n_s/2$) are marked. Inset, optical microscope image of the device. The leads for the top and bottom graphite gates are noted with TG and BG. \textbf{e,} Development of the insulating states with parallel magnetic fields. The top and bottom panels are zoom-in scans corresponding to areas marked by the dashed square in \textbf{d}. \textbf{g,} Conductivity as a function of density along the dashed line marked in \textbf{d} at various temperatures. Inset, Arrhenius plot for the full-filled insulating state ($n_s$) and half-filled insulating state ($n_s/2$) under different parallel magnetic fields.  \textbf{f,} Effective mass measured by temperature dependent quantum oscillations along the same line as \textbf{g}. \textbf{h,} Half-filled insulating gap $\Delta_{n_s/2}$ and quarter-filled insulating gap $\Delta_{n_s/4}$ as a function of parallel magnetic field. The dashed line indicates Zeeman energy with $g=2$. Insets, Schematic diagram of density of states. The gray band represent v$_1$, while the orange color represents spin-polarized half-filled c$_1$ band and red depicts the upper spin band of c$_1$. Black arrows indicate the spin-polarization direction of the corresponding band. The orange and red bands are further split into four bands under strong parallel magnetic fields.}

%\end{center}

\begin{figure}
\includegraphics[width=\textwidth]{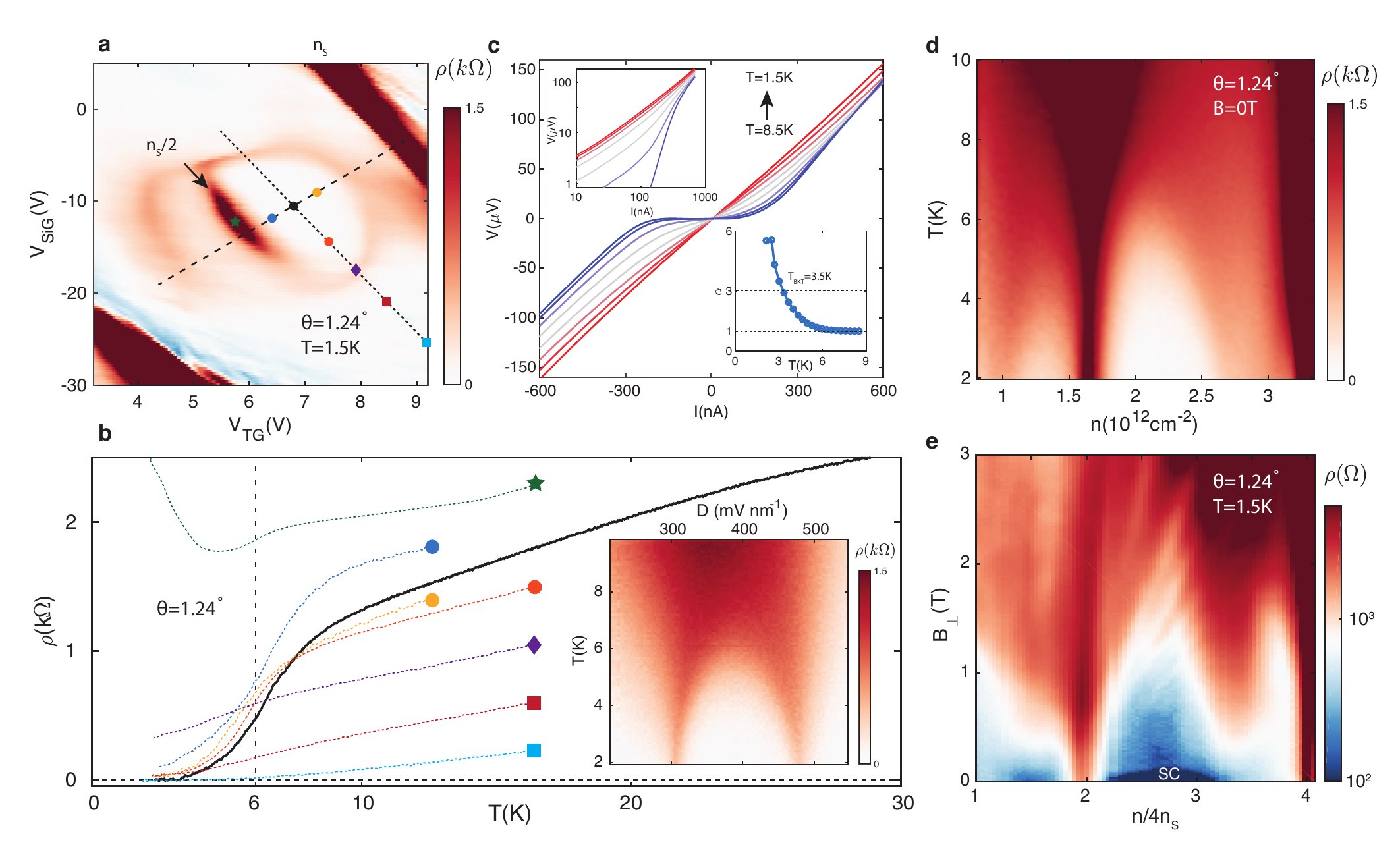}

\caption{\textbf{$\vert$ Superconductivity in $1.24^\circ$ sample.} \textbf{a,} Resistivity map around the half-filled insulator and superconductor. \textbf{b,} Resistivity as a function of temperature at different spot marked by colored symbols in \textbf{a} using the corresponding colors and shapes. Inset, temperature dependence along the constant density line-cut marked by dotted line in \textbf{a}. \textbf{c,} $I$--$V$ curves at optimal doping $n_m$ and displacement field $D_m$ (black circle in Fig. 2\textbf{a}), Top left inset, $I$--$V$ in log scale. Bottom right inset, power $\alpha$ ($V \propto I^\alpha$) as a function of temperature. \textbf{d,} Temperature dependence of the constant displacement field line-cut (dashed line in \textbf{a}). \textbf{e}, Perpendicular magnetic field dependence along the same line in \textbf{d} at $T=1.5$~K.}
\end{figure}

\begin{figure}
\includegraphics[width=\textwidth]{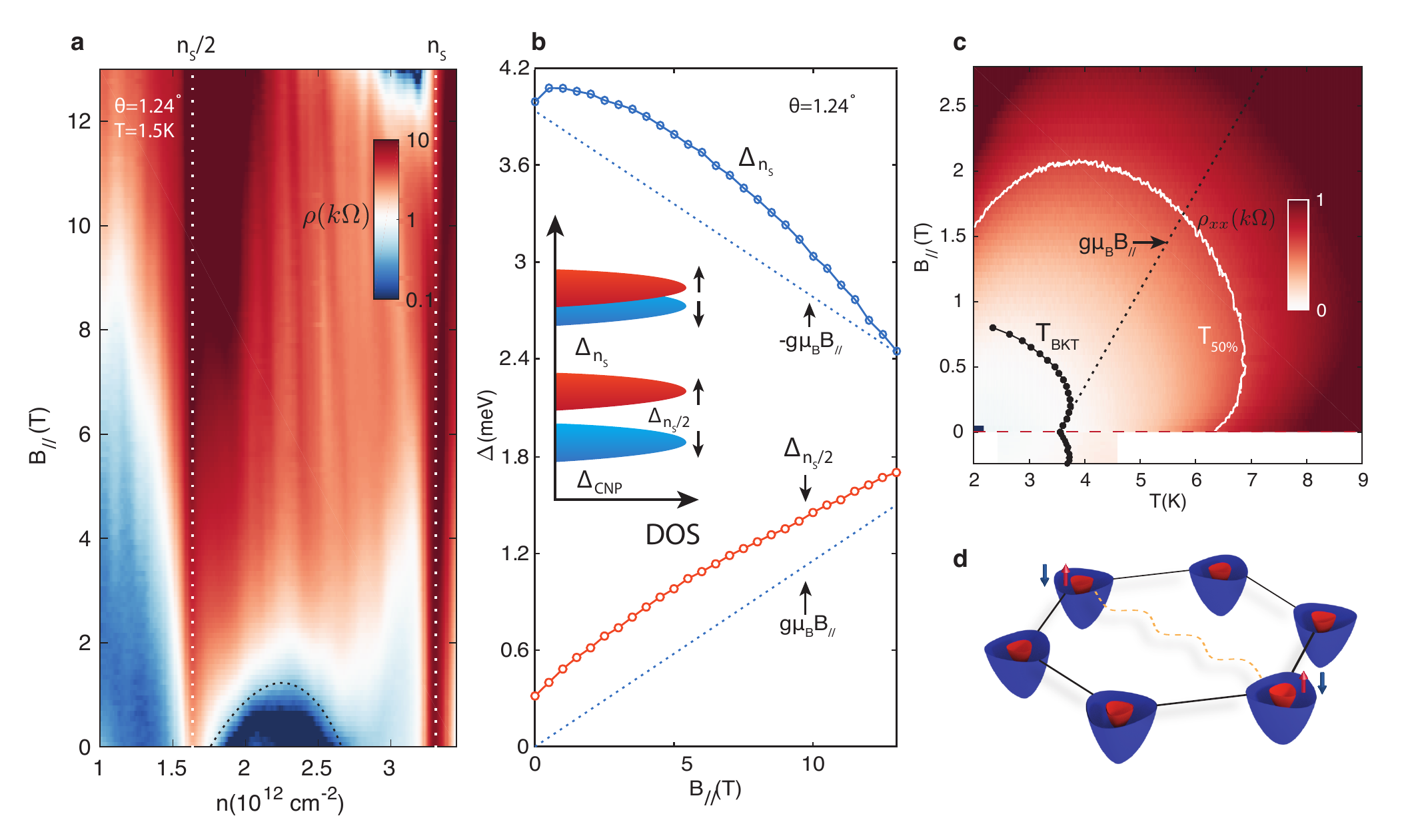}
\caption{\textbf{$\vert$ Spin-polarized superconductivity.} \textbf{a,} Parallel field dependence of resistivity along the same constant displacement field line as Fig. 2\textbf{d} and 2\textbf{e}. The color plot is in log scale, with superconducting region noted by dashed half-circle. \textbf{b,} Half-filling and full-filling gap as a function of parallel magnetic field in the $\theta=1.24^\circ$ device. Dashed lines mark the Zeeman energy with $g=2$ for comparison. \textbf{c,} Resistivity as a function of temperature and parallel magnetic field at the optimal doping $n_m$ and displacement field $D_m$. The BKT temperature is overlaid on the graph (black circles and line). Red dashed line marks zero magnetic field, while black dotted line represents $k_B \Delta T_{BKT}\approx g\mu_B |B_{\parallel}|$. \textbf{d,} Illustration of possible ferromagnetic pairing mechanism. The hexagon represents Brillouin zone of graphene lattice. Pairing (as indicated by wavy dashed line) occurs between electrons in the opposite valleys with the same spin.}
\end{figure}

\end{document}

\maketitle

\subsection*{S1 Theoretical band structure of twisted double bilayer graphene} 

\begin{figure}[H]
 \centering
\includegraphics[width=0.83\textwidth]{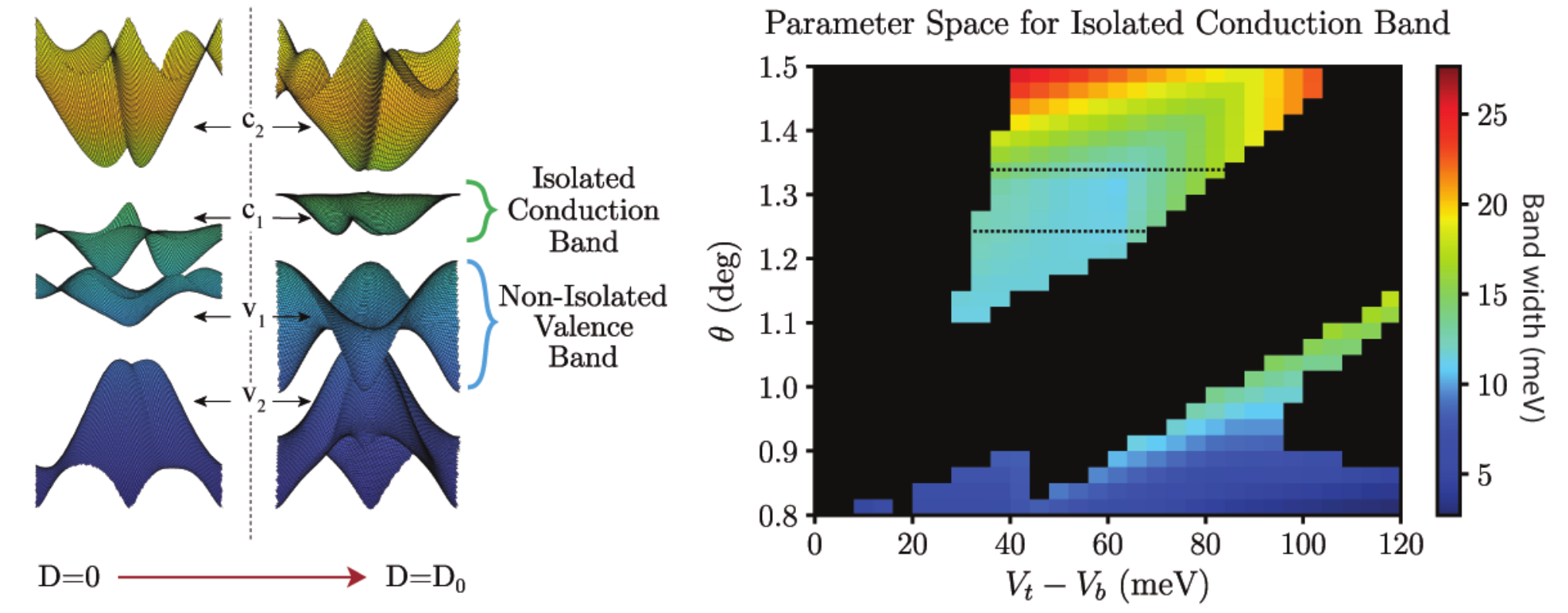}
\includegraphics[width=0.83\textwidth]{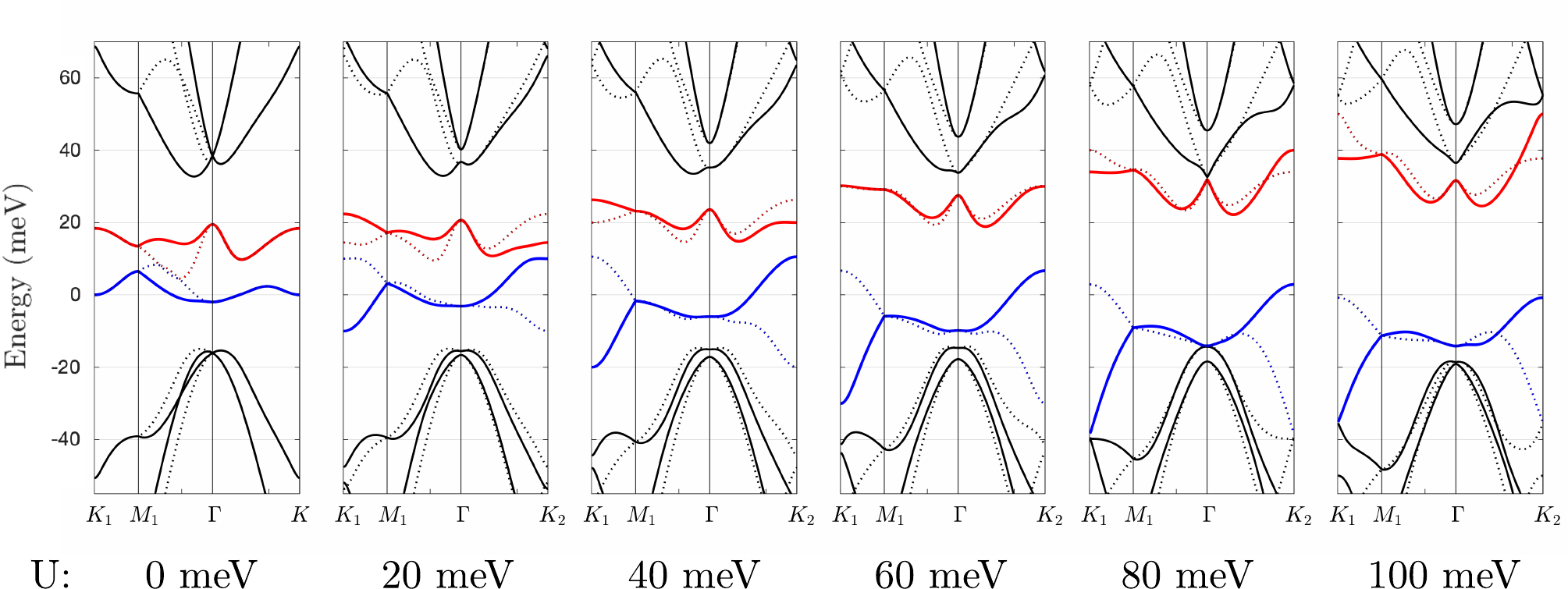}
 \caption{(Top Left), calculated band structure of TBBG at zero displacement field and optimal displacement field $D_0$ for the isolated flat band. (Top Right), calculated parameter space for isolated conduction band (x-axis is onsite potential difference $U = V_t - V_b$ between the top and bottom bilayer graphene, y-axis is twist angle). Color represent bandwidth of the first conduction band $c_1$ (meV). In the colored parameter space, c$_1$ is isolated from the remote band and valance bands. Dotted line represents cuts at $\theta=1.24,1.33^\circ$, corresponding to Fig. S~\ref{FS2}. (Bottom), band structures at $\theta = 1.24^\circ$ for increasing $U$. Color: red for c$_1$ and blue for v$_1$. Dotted lines are for bands in the other valley.  }{\label{FS1}}
\end{figure}

Fig. S$\ref{FS1}$ shows the evolution of the moir\'e band structure with increasing $D$. The band structure of TDBG with Bernal-stacked bilayers is obtained as the following. In TDBG, each bilayer graphene has a tight-binding Bloch Hamiltonian at a momentum $\bk$ given by

\begin{equation}
H(\bk)=\mqty( U_{1}+\Delta & - \gamma_0 f(\bk) & \gamma_4 f^*(\bk) & \gamma_1 \,\, \\
		 - \gamma_0 f^*(\bk) & U_{1} & \gamma_3 f(\bk) & \gamma_4 f^*(\bk) \\
        \gamma_4 f(\bk) & \gamma_3 f^*(\bk) & U_{2} & -\gamma_0 f(\bk) \\
        \gamma_1& \gamma_4 f(\bk) & - \gamma_0 f^*(\bk) & U_{2}+\Delta ),
\end{equation}
which is labelled in the order of  $A_\textrm{1}$, $B_\textrm{1}$, $A_\textrm{2}$, $B_\textrm{2}$ sites of the top (1) and the bottom (2) Bernal stacked bilayer graphene. $f(\bk) \equiv \sum_l e^{i \bk \cdot \delta_l}$,
where $\delta_1 = a(0,1)$, $\delta_2 = a(\sqrt{3}/2,-1/2)$, $\delta_3 = a(-\sqrt{3}/2,-1/2)$ with $a=1.42 \textrm{ \AA}$. In particular, the electrostatic energy difference $U$ between top and bottom-most layers is an important tuning parameter controlled by $D$. With this Hamiltonian, one can follow the continuum model approach in Ref.~\cite{MacDonald2011} to calculate the moir\'e band structure.

In the numerical simulation, we use phenomenological parameters 
\begin{equation}
    (\gamma_0, \gamma_1, \gamma_3, \gamma_4, \Delta) = (2610,361,283,138,15) \textrm{ meV}
\end{equation} obtained from Ref.~\cite{Jeil2014}. Compared to TBG, TDBG has additional parameters $\gamma_1$, $\gamma_3$ (trigonal warping), $\gamma_4$ (particle-hole asymmetry) and $\Delta$, in addition to $\gamma_0$ (nearest-neighbor hopping). Here, $\gamma_1$ and $\Delta$ are the inter-layer hopping and the on-site energy at A-B stacked sites where the $A$-site of the first layer ($A_1$) sits on top of the $B$-site of the second layer ($B_2$), respectively. Although these parameters are much smaller than $\gamma_0$, they are important to understand the experimental data. Particularly, for vanishing $U$, a finite value of $\gamma_3$ yields a larger bandwidth and overlap between c$_1$ and v$_1$. This is why the system is metallic at CNP and there is no magic-angle condition at $D=0$. Furthermore, $\gamma_4$ and $\Delta$ give rise to the electron-hole asymmetry. Due to these terms, the bandwidth of v$_1$ is much larger than that of c$_1$, resulting in smaller band isolation for v$_1$. See Fig.~S\ref{FS1} for the comparison. For the moir\'e hopping parameters, $(w_0,w_1) = (0.08,0.1)$~eV is used to account for the relaxation effect described by Ref.~\cite{Moon2013}. The relaxation increases the gap between c$_1$ and c$_2$ (v$_1$ and v$_2$), stabilizing the insulating states for the range of $D$ at $\nu = \pm 4$ fillings.  For more details, See Ref.~\cite{TDBGtheory2019}.

\subsection*{S2 Experimental displacement field ranges for isolated bands, correlated insulators and superconductivity} 
\begin{figure}[H]
 \centering
\includegraphics[scale=0.67]{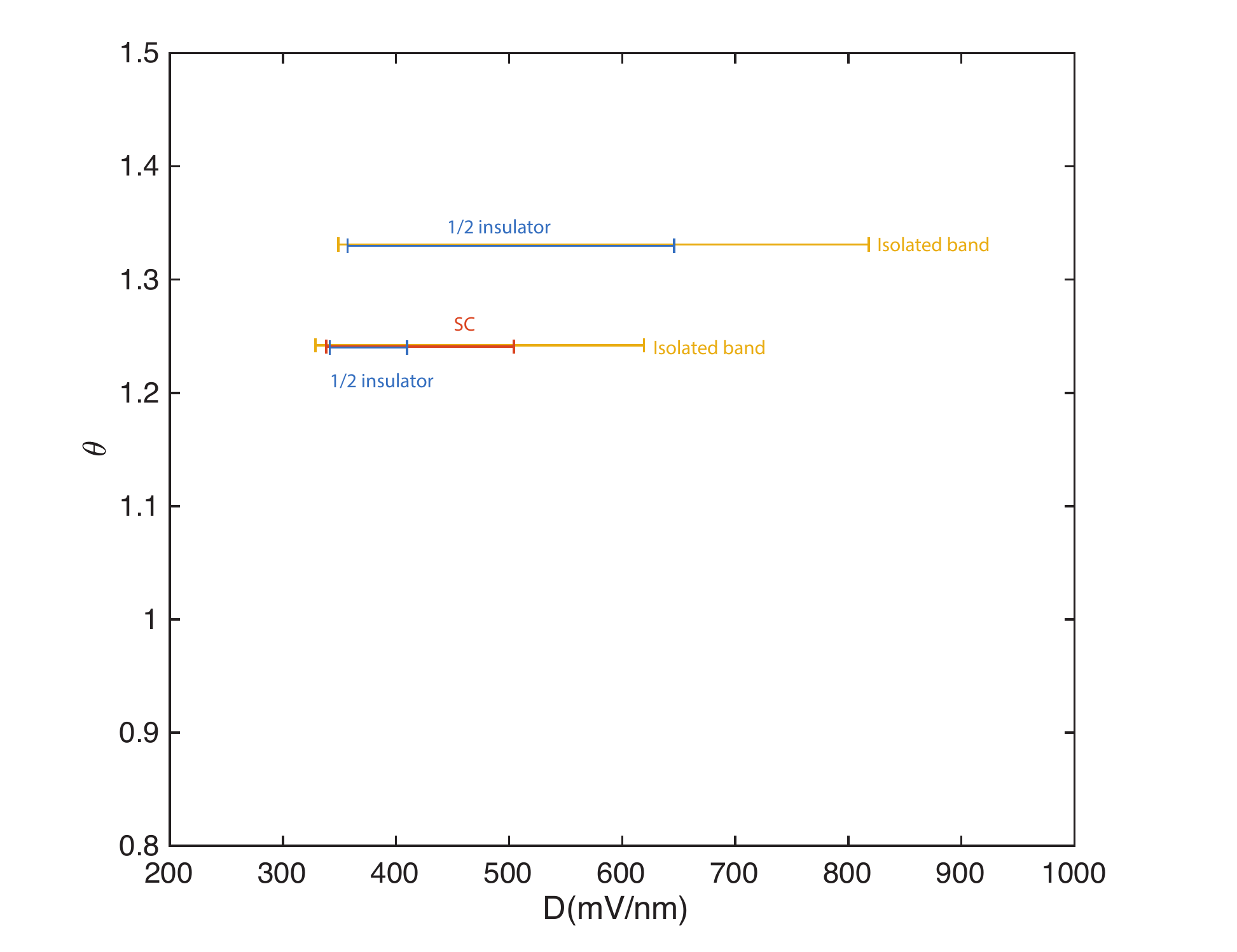}
 \caption{Experimentally observed displacement field ranges for different features observed in the two samples with different twist angles.}{\label{FS2}}

\end{figure}

Fig. S\ref{FS2} shows the experimental displacement field ranges where the correlated insulating states at half-filling ($\Delta_{n_s/2}>0$), superconductivity and the isolated conduction band are observed. At a certain displacement field, isolated band refers to the coexistence of a bandgap between c$_1$ and c$_2$ ($\Delta_{n_s}>0$) as well as a bandgap between c$_1$ and v$_1$ ($\Delta_{\textrm{CNP}}>0$). As we can see in Fig. S\ref{FS2}, the onsets of the correlated insulators and the superconductivity occur at the same displacement field where the band isolation becomes appreciable.. However, correlated behaviors disappear before the band isolation vanishes at higher displacement field. The above observations suggest the appearance of the correlated states requires both isolated and narrow single particle band.

\subsection*{S3 Additional data for $\theta=1.33^\circ$ sample}

\subsubsection*{S3.1 Temperature dependence of resistance around half-filling}
\begin{figure}[H]
 \centering
\includegraphics[scale=0.8]{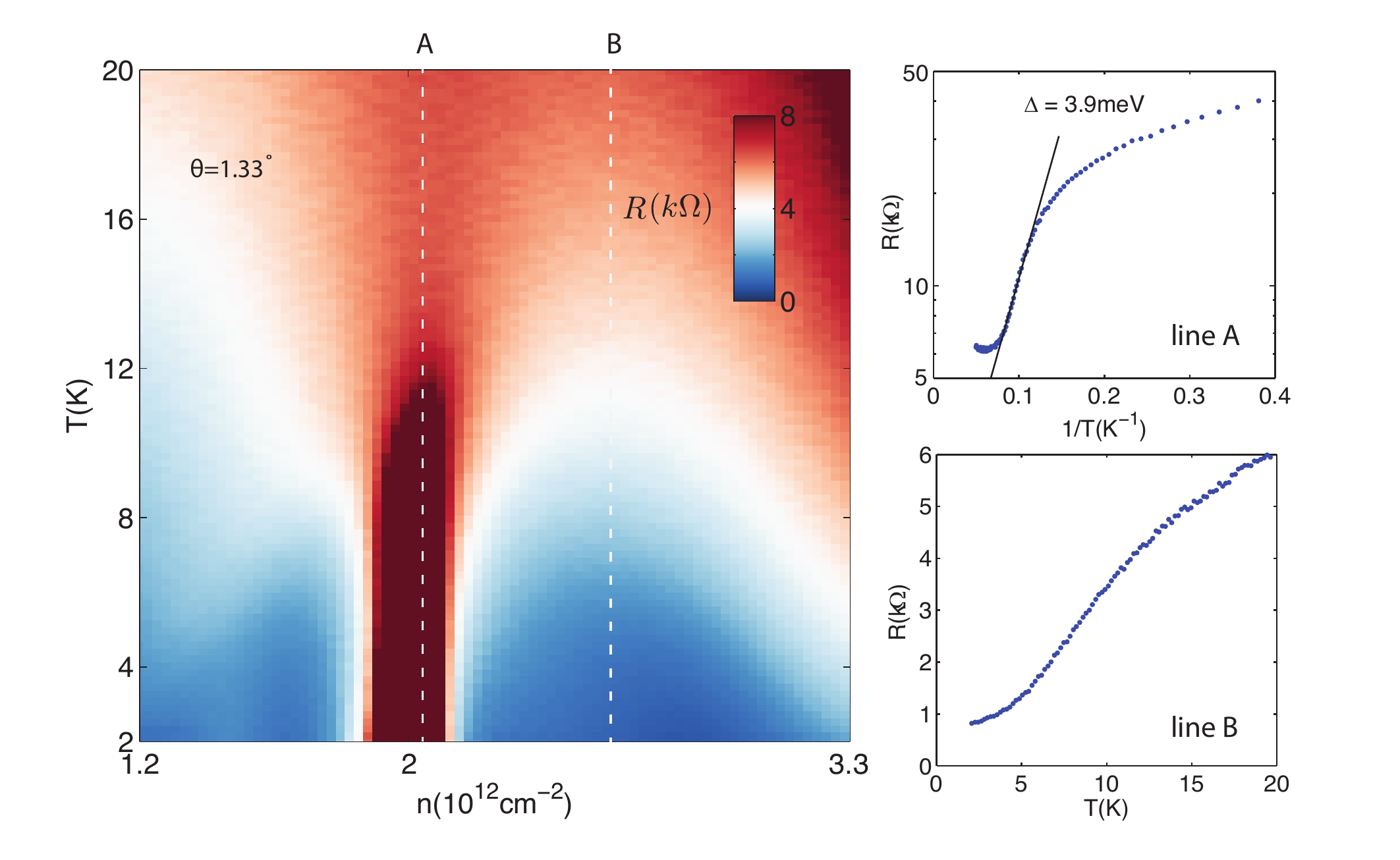}
 \caption{Resistance as a function of density and temperature in $\theta=1.33^\circ$ device. Line A cuts through the half-filled insulator $n=n_s/2$. Line B cuts through a low resistance area next to the half-filled insulator, which shows that the resistance drops with decreasing temperature but does not go to zero.}{\label{FS3}}

\end{figure}

Fig. S\ref{FS3} shows the temperature dependence of resistance R(T) at different densities measured in $\theta=1.33^{\circ}$ device. In particular, we pick two representative R(T)s corresponding to the most insulating regime (line A) and the most conducting regime (line B) at low temperatures. The energy gap $\Delta_{n_s/2}\approx 3.9$~meV is obtained from the activating behavior along line A. For the most conducting state (along line B), we have not observed superconducting transition down to 300~mK. Instead, metallic behavior (resistance becomes smaller with decreasing T) persists.

\subsubsection*{S3.2 Landau Fan diagram}

\begin{figure}[H]
 \centering
\includegraphics[scale=0.78]{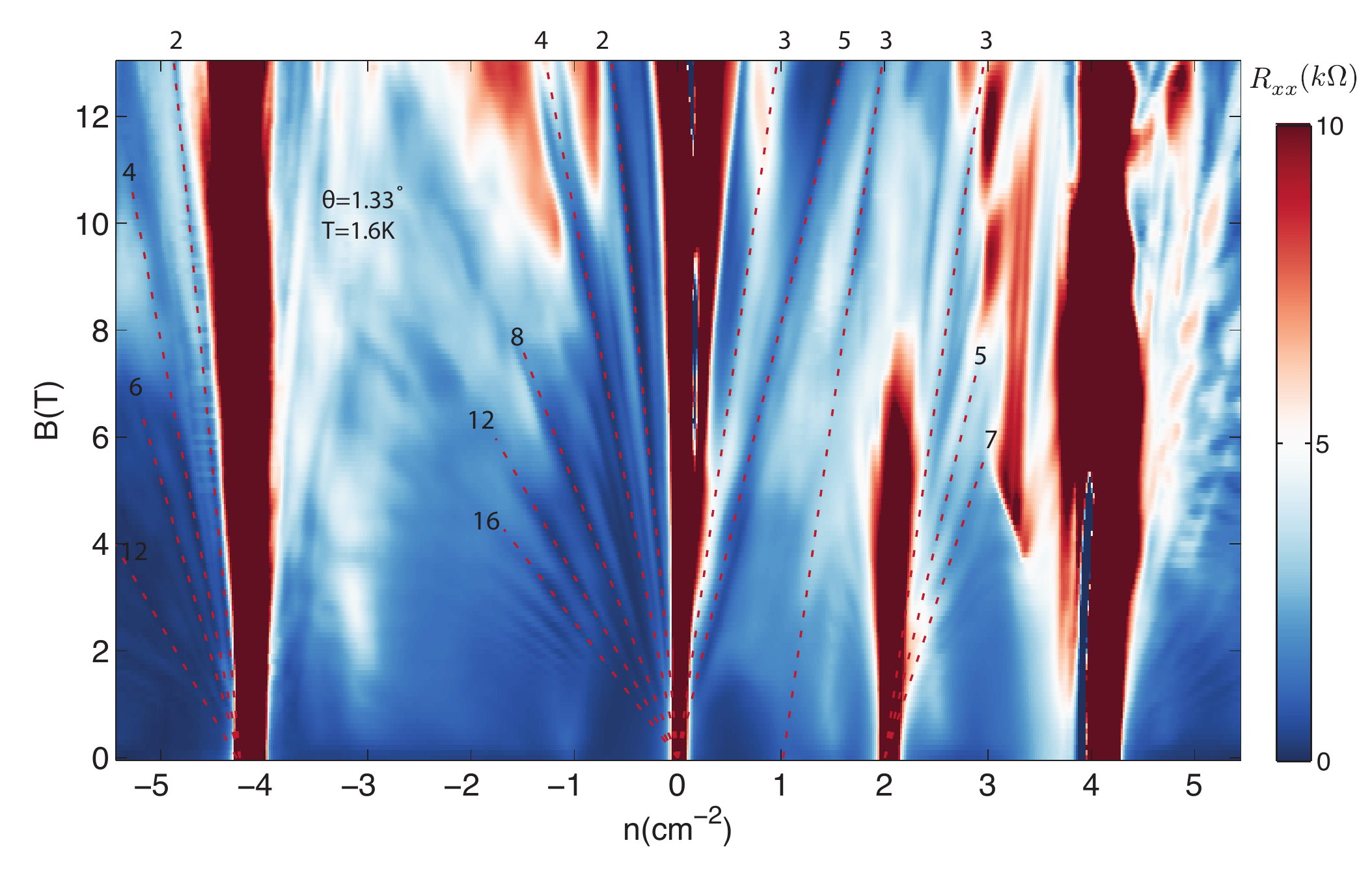}
 \caption{The Landau fan diagram of longitudinal resistance as a function of density and perpendicular magnetic field. The numbers in the plot indicate Landau level filling fractions corresponding to the lines next to them}{\label{FS4}}

\end{figure}

Under a perpendicular magnetic field, clear fans can be identified coming from the charge neutral point, full-fillings $\pm n_s$ ($ \approx \pm4.1\times10^{12}~\textrm{cm}^{-2}$) as well as half-filling $n_s/2$ on the electron-doped side. The Landau fan from the charge neutral point exhibits well-developed quantum Hall (QH) states with four fold degeneracy on the valence band side under low magnetic fields, and subsequently develops the full degeneracy-lifted QH states under higher magnetic fields. The Landau fans on the conduction band side (n$>$0) is highly unusual. The fan from the charge neutral point exhibits only an sequence of odd filling fractions $\nu=$ 3 and 5. While the fan coming from the correlated insulating state at half-filling shows a degeneracy of two, consistent with the picture of a spin-polarized half-filled band, the sequence is also of odd numbers $\nu=$ 3, 5 and 7. There is also one QH state $\nu=$3 projected down to quarter-filled conduction band, suggesting spin-valley polarization and possibly non-zero Chern numbers associated with this band. Further study is required to clarify these experimental observations. Above a perpendicular magnetic field of 7~T, the half-filled insulator disappears, presumably due to the orbital effect of the perpendicular magnetic field.

\subsubsection*{S3.3 Effective mass calculation}

We calculate the effective cyclotron mass from temperature dependent magnetoresistance (SdH) oscillations. The cyclotron mass is a measure of density of state and thus directly related to the Landau level separation (cyclotron gap) under a given magnetic field. As temperature rises, SdH oscillation amplitude is reduced following 
\begin{equation}
    \Delta R \propto \chi/ \sinh(\chi), \textrm{in which } \chi=\frac{2\pi^2 k  m^*}{\hbar e} \frac{T}{B}. 
\end{equation}

\begin{figure}[H]
 \centering
\includegraphics[scale=0.4]{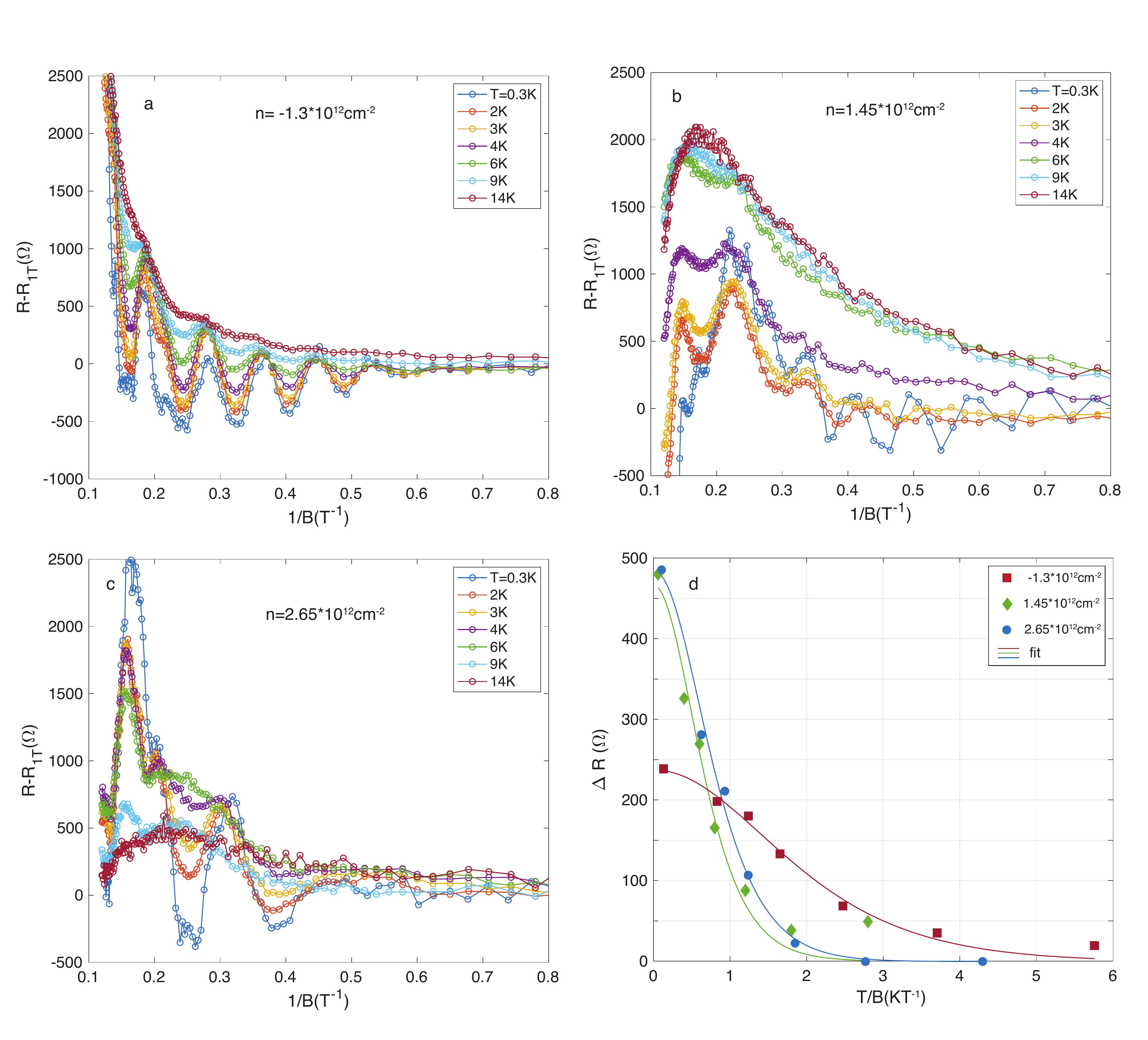}
 \caption{Effective mass calculation. \textbf{a}-\textbf{c}, temperature dependent SdH oscillations at a few representative density points: \textbf{a}, in the valance band, \textbf{b}, in the conduction band below half-filling, \textbf{c}, in the conduction band above half-filling. \textbf{d}, extracted oscillation amplitudes as a function of $T/B$ for the density configuration shown in \textbf{a}-\textbf{c} and corresponding fitting curves.}{\label{EffMass}}
 
 \end{figure}

We measured SdH oscillation at all densities between filling factor $n/n_S$ = -1 and 1, at T=~0.3, 2, 3, 4, 6, 9, 14~K (example: Fig. S\ref{EffMass}a-c). We then extracted the oscillation amplitudes and plot them as a function of $T/B$. Fitting $\Delta R (T/B)$ with the above formula with $m^*$ being the only fitting parameter, we obtain the effective cyclotron mass shown in main text.

\subsubsection*{S3.4 Perpendicular magnetic field effect on the half-filled insulating state}

\begin{figure}[H]
 \centering
\includegraphics[scale=0.75]{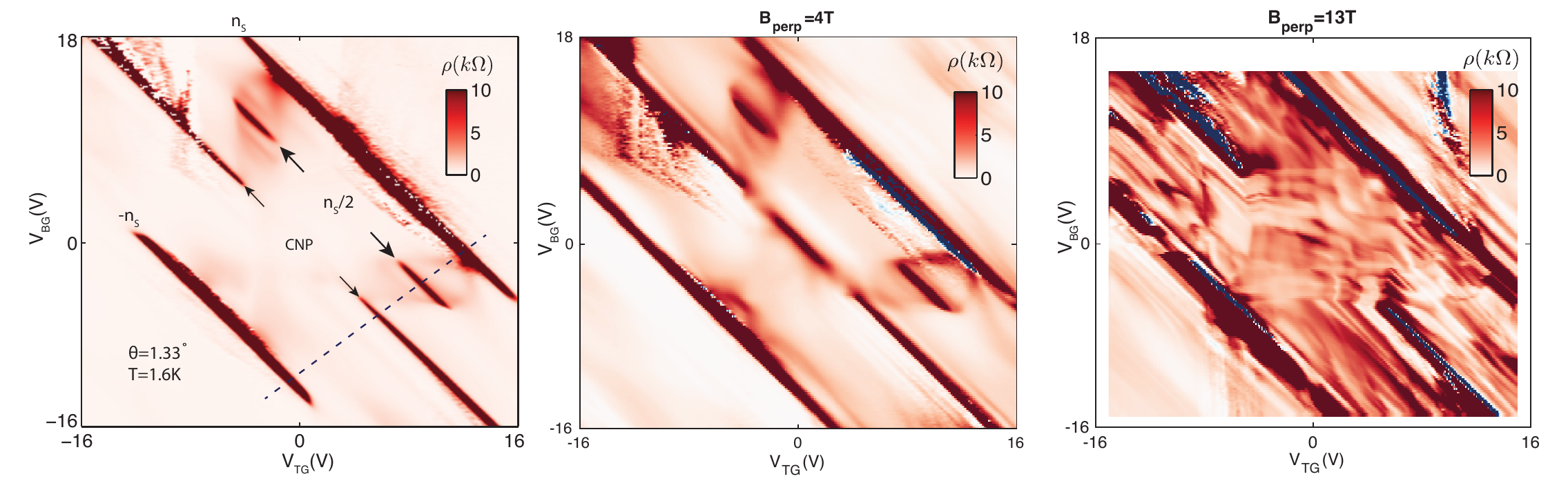}
 \caption{Resistivity as a function of top and bottom gate voltages under perpendicular magnetic field of 0~T (left), 4~T (middle) and 13~T (right). Note that the position of the half-filled insulating state remains unchanged under $B_{\perp}\approx 4$~T, but then it is taken over by Landau level formation in the high magnetic field regime ($B_{\perp}\approx 13$~T).}{\label{PerpB}}
 
 \end{figure}

As shown in Fig. S\ref{FS4}, the half-filled insulator disappears above a perpendicular magnetic field of 7~T. However, below this critical field, the displacement field range where the half-filled insulator resides remain the same as zero magnetic field, as shown in Fig. S\ref{PerpB}.

\subsection*{S4 Large twist angle $\theta=2^\circ$ sample}
\begin{figure}[H]
 \centering
\includegraphics[scale=0.75]{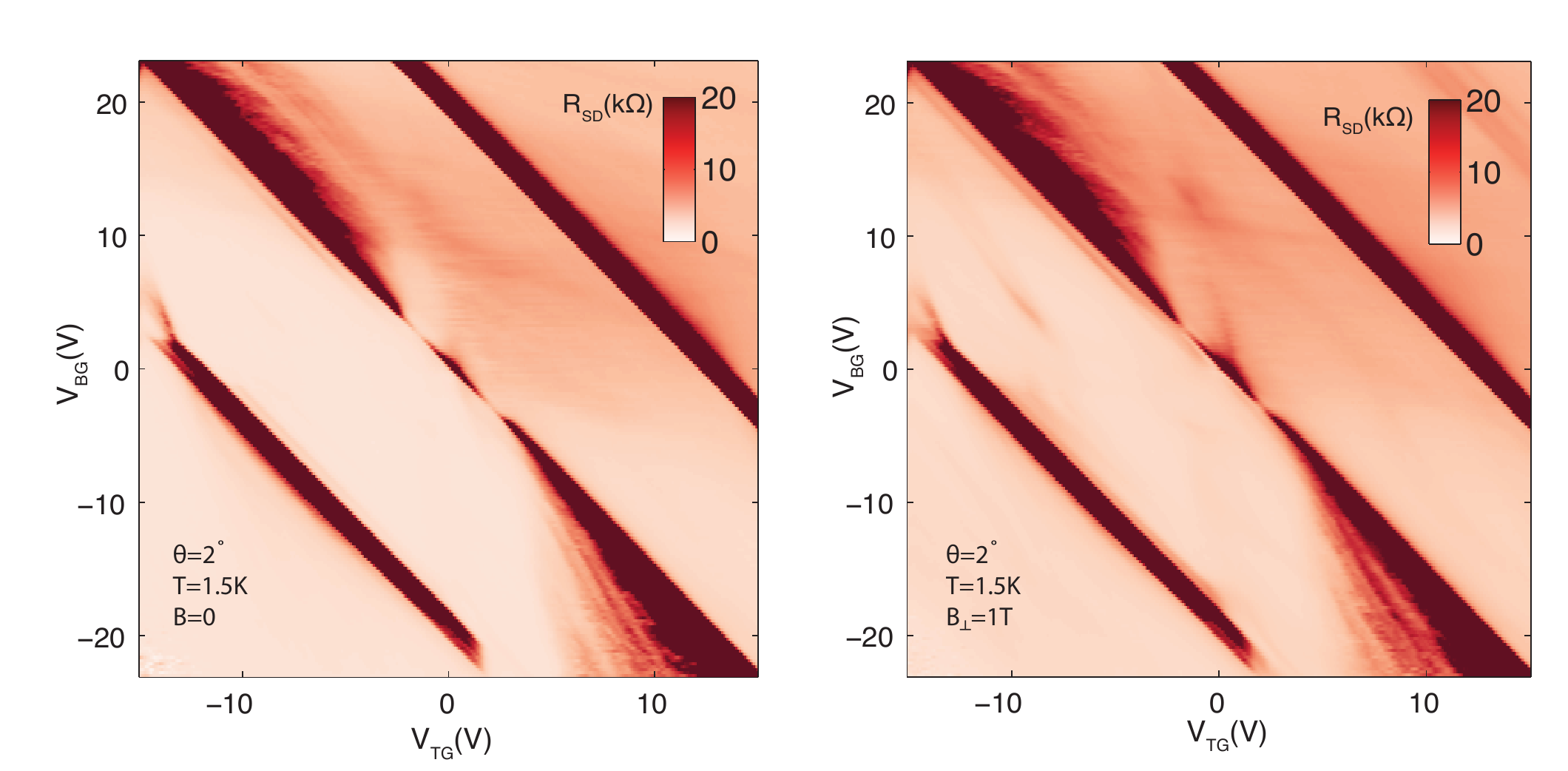}
 \caption{Left, dual gates dependence of two terminal resistance under zero magnetic field. Right, the same measurement as left but under a perpendicular magnetic field of 1~T.}{\label{FS5}}

\end{figure}

In the samples with larger twist angles ($\theta=1.9^\circ$ and $2^\circ$), we do not observe signs of correlated insulators nor superconductivity (Fig. S\ref{FS5}), despite that isolated conduction and valance band persist over most of the displacement field range experimentally accessible. At zero displacement field, there is a gap between conduction and valance band, which closes and reopens as displacement field increases. This gap closing and openning is explained by the numerical calculation as well~\cite{TDBGtheory2019}. The absence of correlated insulator states is presumably due to the absence of flat bands (numerically calculated bandwidth is about $80$~meV). Under a small perpendicular magnetic field ($B_{\perp}$ = 1~T), a slight increase of resistance appears near half-filling of the valance band. However, this 
sample never exhibits a fully insulating behavior at half-filling under our experimental conditions.

\subsection*{S5 Additional data for $\theta=1.24^\circ$ sample}

\subsubsection*{S5.1 Resistivity over full gate range}
\begin{figure}[H]
 \centering
\includegraphics[scale=0.7]{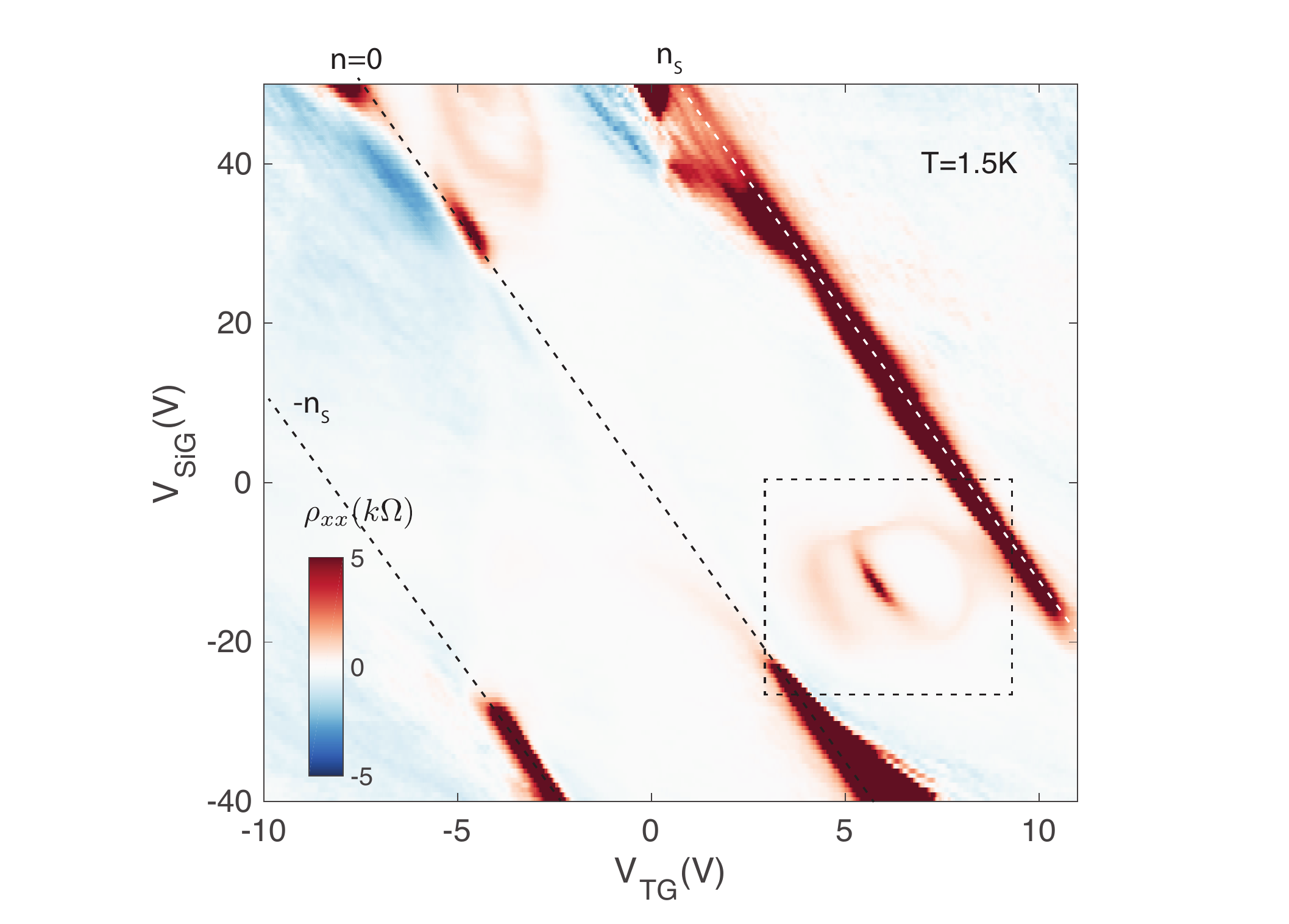}
 \caption{Resistivity over a large range of gate voltages in the $\theta=1.24^\circ$ sample. The dashed rectangle marks the zoomed-in scan shown in the main figure Fig.2a}{\label{FS6}}

\end{figure}

Fig. S\ref{FS6} shows resistivity measured over a large gate voltage range for the $\theta=1.24^\circ$ sample.  The silicon substrate is used as the back gate in this device. The lack of symmetry between the top graphite gate and silicon backgate is likely caused by less transparent contacts. A large contact resistance can occur due to the formation of p-n junctions at the leads near the contacts when the top gate is biased towards large negative voltages. Poor quality electrical contacts can be inferred from the negative resistivity with large noise when the top gate voltage is below -4~V. However, we emphasize that the contact transparency in this device remains excellent in the gate range we focused on in the main text (Fig. 2 and Fig. 3) (marked by dashed square region in Fig. S8), since both channel and contact regions remain in the same polarity of carriers. 

\subsubsection*{S5.2 Transport behaviors outside the superconducting ring}

\begin{figure}[H]
 \centering
\includegraphics[scale=0.62]{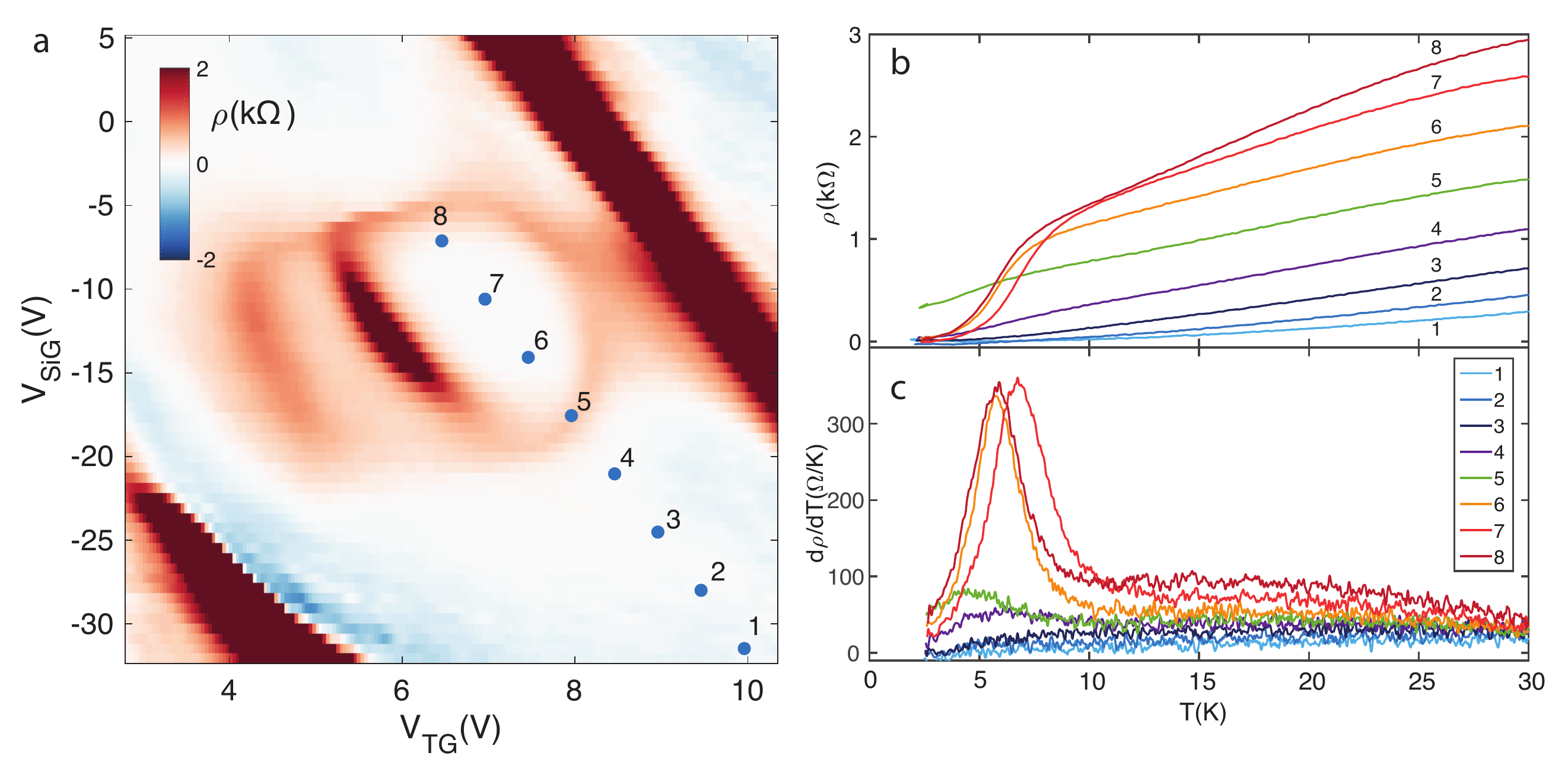}
 \caption{Temperature dependence of resistivity at different gate configurations. \textbf{a,} dual gate map showing the gate voltage configurations of curve 1-8. \textbf{b} and \textbf{c,} $\rho$ and d$\rho$/dT as a function of temperature measured at the spots marked by 1-8 in left plot. Curve 5 is on the boundary of the ring structure, while 1-4 are outside and 6-8 are inside.}{\label{FS17}}
\end{figure}

The temperature dependence of resistivity outside the superconducting ring does not exhibit transitioning behaviors (curve 1-4 in Fig. S\ref{FS17}b\&c). Unlike the sharp superconducting transitions observed inside the ring (curve 6-8 in Fig. S\ref{FS17}b\&c), the resistivity outside the ring increases roughly linearly with increasing temperatures. As the displacement field moves away from the ring, the temperature dependence become even weaker (curve 1 in Fig. S\ref{FS17}b\&c). One possible cause of this low-resistivity behavior is ballistic transport of carriers. As the displacement field deviates from the flat band condition, the band becomes more dispersive with a finite group velocity, and thus ballistic transport could occur similar to that in a single bilayer graphene at low temperatures \cite{Wang_1D}. In the multi-terminal ballistic transport regime, one may expect very low (or even negative) voltage readings between the voltage probes depending on the mesoscopic coupling between the contacts. As temperature rises, electron-phonon scattering causes the resistivity to increase gradually. 

\begin{figure}[H]
 \centering
\includegraphics[scale=0.55]{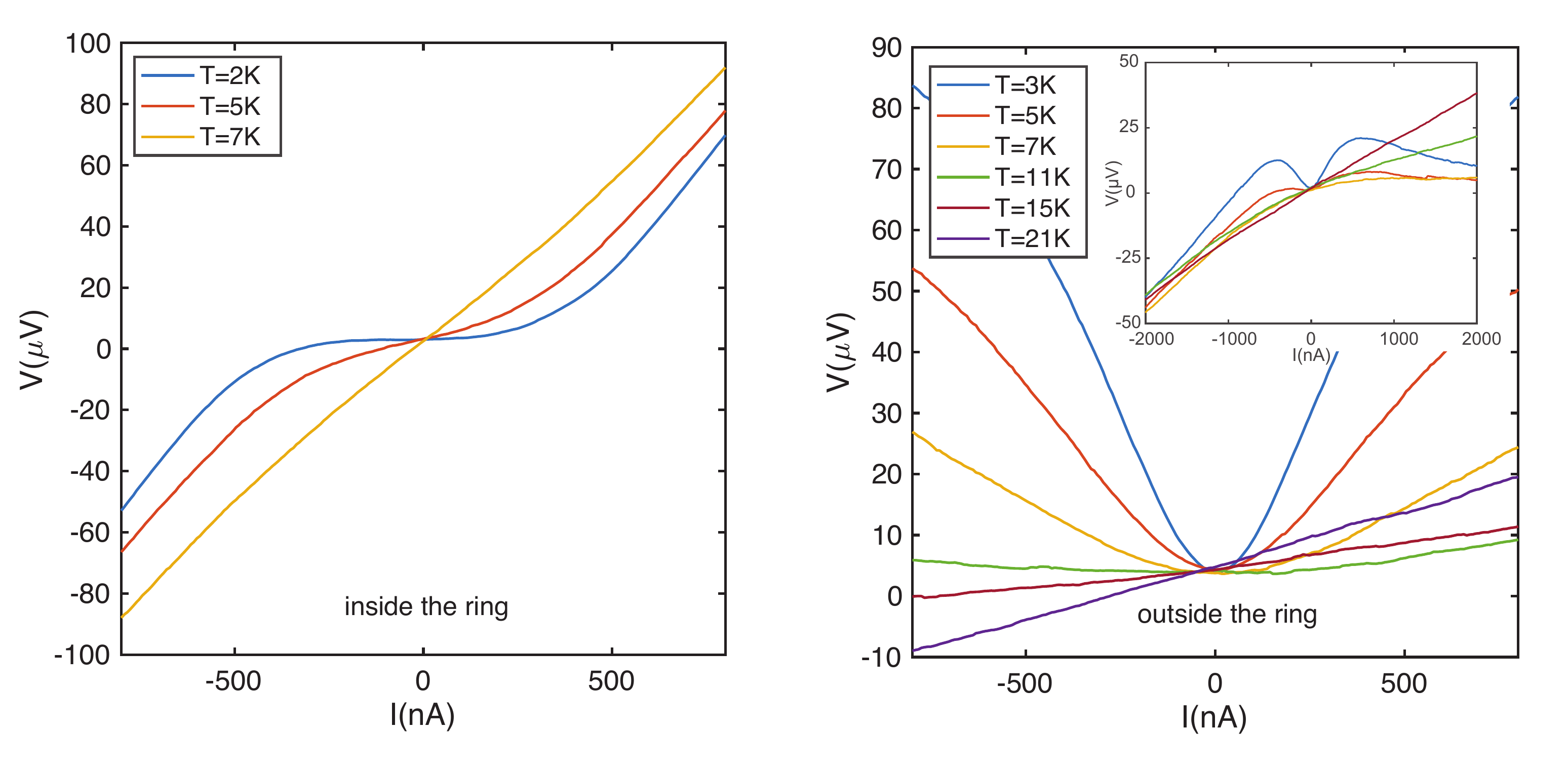}
 \caption{$I$--$V$ characteristics inside (left) and outside (right) the ring shown in Fig. S\ref{FS17}. Inset on the right shows $I$--$V$ characteristics with large biases at a point outside the ring.  }{\label{FS18}}
\end{figure}

The contrast between the gate regimes inside the ring (superconducting) and outside the ring (non-superconducting) can be further demonstrated from the finite bias $I$--$V$ measurements. In Fig. S\ref{FS18}, data in the left panel are taken inside the ring, regime of the gate configuration in Fig. S9 where we observe superconductivity. We observe a typical superconducting $I$--$V$ behavior from which we extract BKT transition temperature. However, outside the ring, the four terminal measurement often exhibits very strong non-linear rectifying behaviors sensitively dependent on gate voltage changes at low temperatures (see right panel of Fig. S10). Such behaviors become more regular linear curves at higher temperatures. While further experimental investigations are necessary, the observed low temperature behaviors are consistent with mesoscopic transport in ballistic conductors.

\subsubsection*{S5.3 Fan diagram and the critical field for the superconductivity under perpendicular magnetic field}

\begin{figure}[H]
 \centering
\includegraphics[scale=0.8]{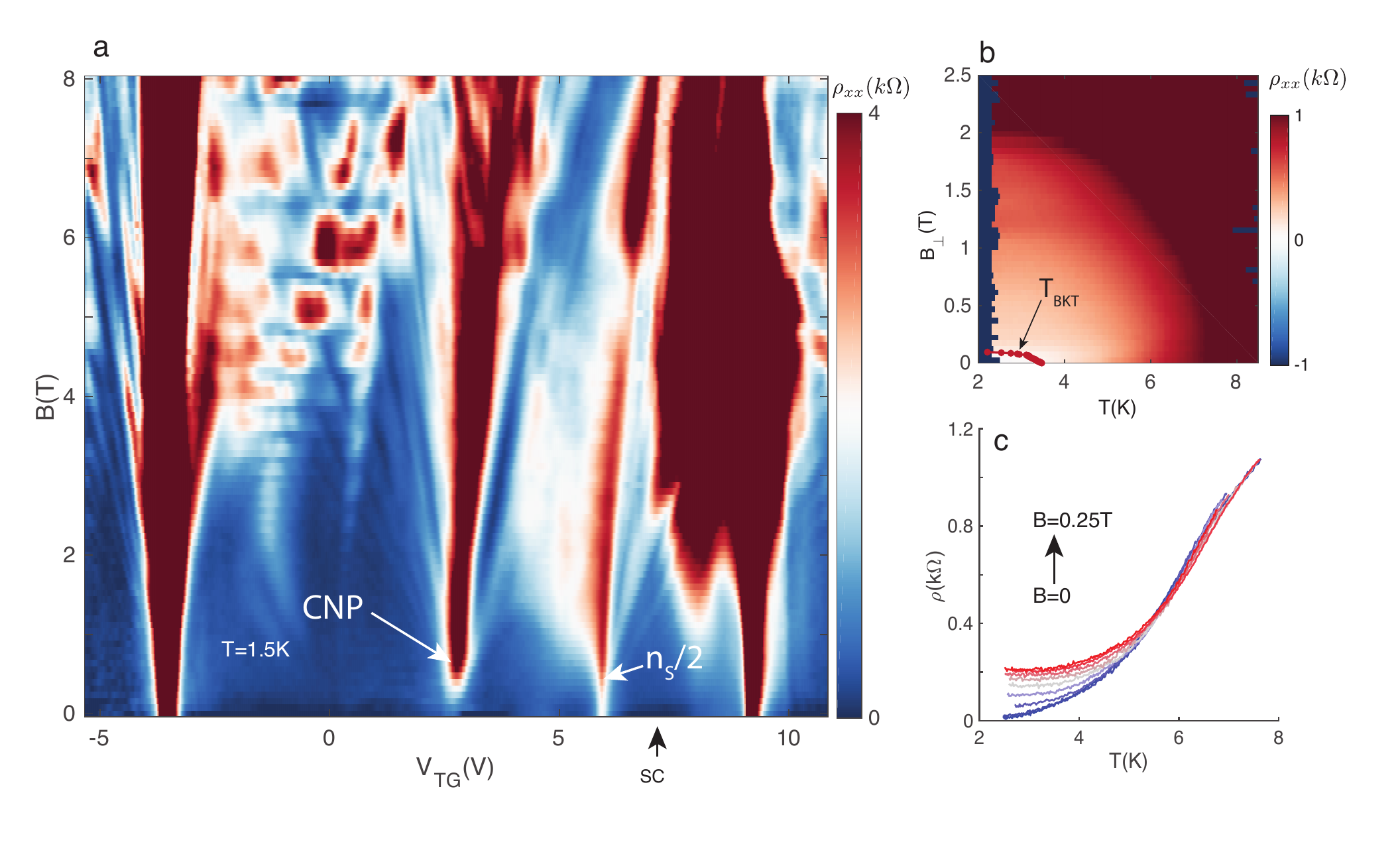}
 \caption{Fan diagram and critical perpendicular magnetic field. \textbf{a}, Fan diagram of the $\theta=1.24^\circ$ sample. The clear Hofstadter butterfly features are used to obtain the twisted angle in this sample. \textbf{b}, perpendicular magnetic field dependence of the superconducting transition at the optimal doping and displacement field ($n_m$, $D_m$). The red curve overlapped on the color plot shows the BKT transition temperatures obtained from $I$--$V$ characteristics. \textbf{c}, resistivity vs. temperature at different perpendicular fields between zero and 0.25~T.}{\label{FS8}}
\end{figure}

Fig. S\ref{FS8} shows perpendicular field dependence of the $\theta=1.24^\circ$ sample. According to BKT transition temperature, the perpendicular critical field is $\approx 0.1$~T. The Landau level degeneracy of fan diagram coming from $n_s/2$ is mostly four (the sequence observed is 6, 12, 16, 20 and 24). 

\subsubsection*{S5.4 Superconductor-insulator transition induced by parallel magnetic fields}
\begin{figure}[H]
 \centering
\includegraphics[scale=0.9]{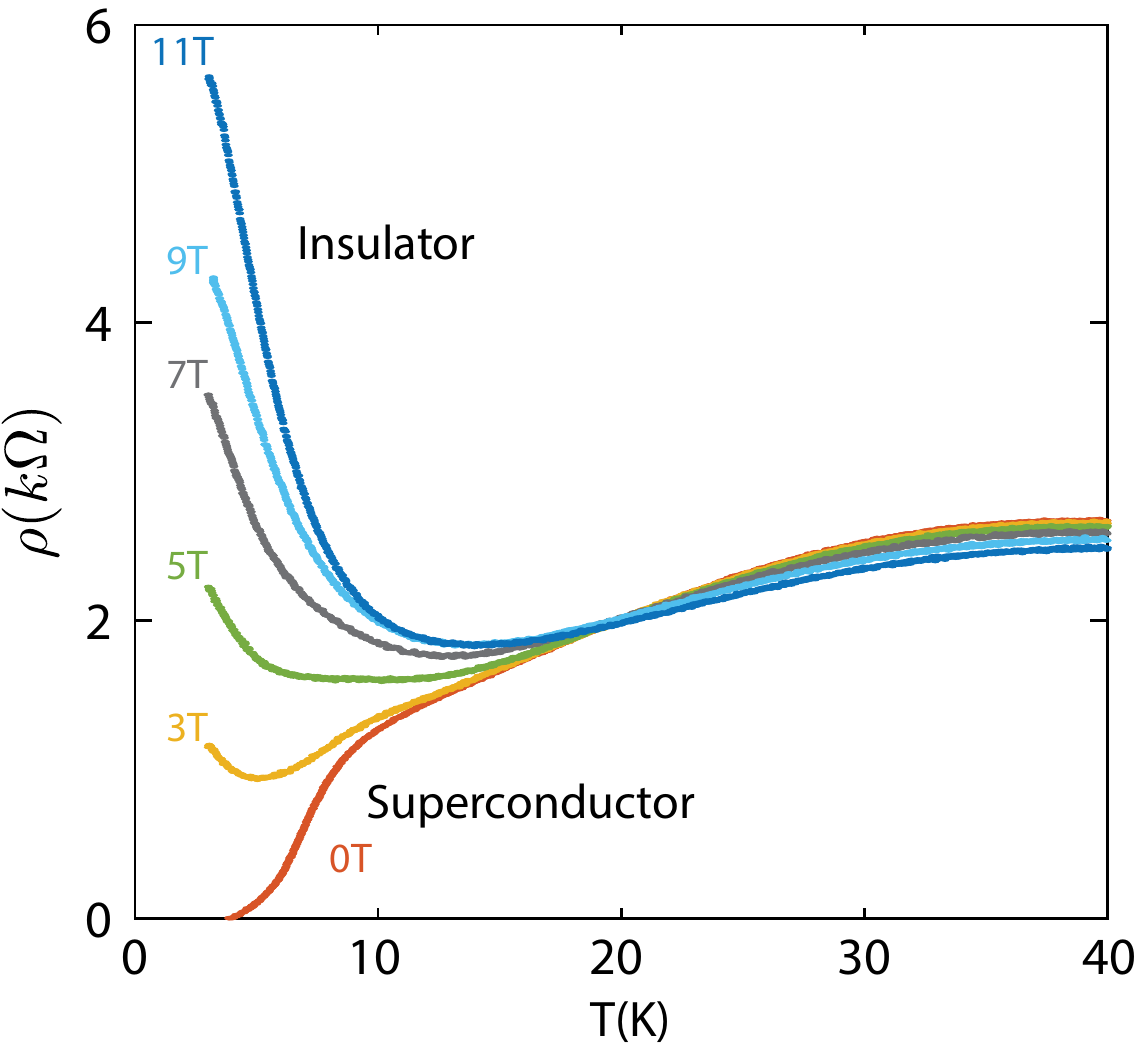}
 \caption{Temperature dependence of the resistivity at optimal doping $n_m$ and displacement field $D_m$ under different parallel magnetic fields}{\label{FS19}}

\end{figure}

Fig. S\ref{FS19} shows the temperature dependence of resistivity in the superconducting phase at optimal doping $n_m$ and displacement field $D_m$ under different parallel magnetic fields $B_{\parallel}$. At zero $B_{\parallel}$, we can identify the superconducting transition. However, as $B_{\parallel}$ increases, the low-temperature behavior of resistivity gradually becomes that of an insulator, with $d\rho/dT <0$ in the low temperature limit. At $B_{\parallel}$=13~T, the low-temperature resistivity is as high as $5.6~k\Omega$. These behaviors suggest a possible superconductor-insulator transition induced by parallel magnetic field.

\begin{figure}[H]
 \centering
\includegraphics[scale=0.8]{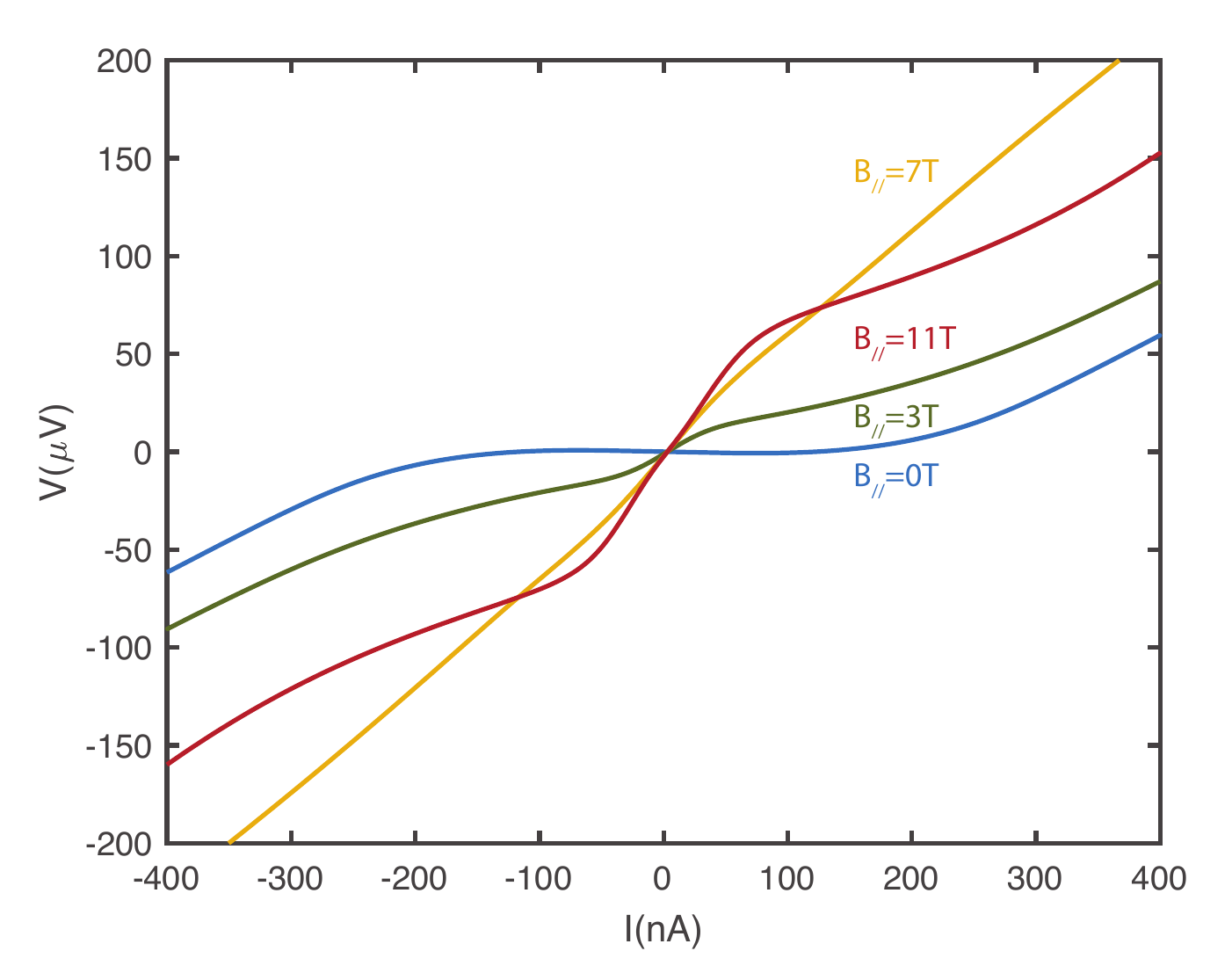}
 \caption{$I$-$V$ characteristic at optimal doping $n_m$ and displacement field $D_m$ under different parallel magnetic fields}{\label{FS11}}

\end{figure}

Furthermore, Fig. S\ref{FS11} shows $I$-$V$ curves at the same doping and displacement field as Fig. S\ref{FS19}. As parallel field increases, the critical current goes to zero and $I$-$V$ becomes Ohmic. Under even higher parallel fields, however, an insulating $I$-$V$ becomes prominent, with differential resistance decreasing as current increases.

\subsubsection*{S5.5 Additional evidences of the enhancement of the superconductivity due to parallel magnetic field}
\begin{figure}[H]
 \centering
\includegraphics[scale=0.8]{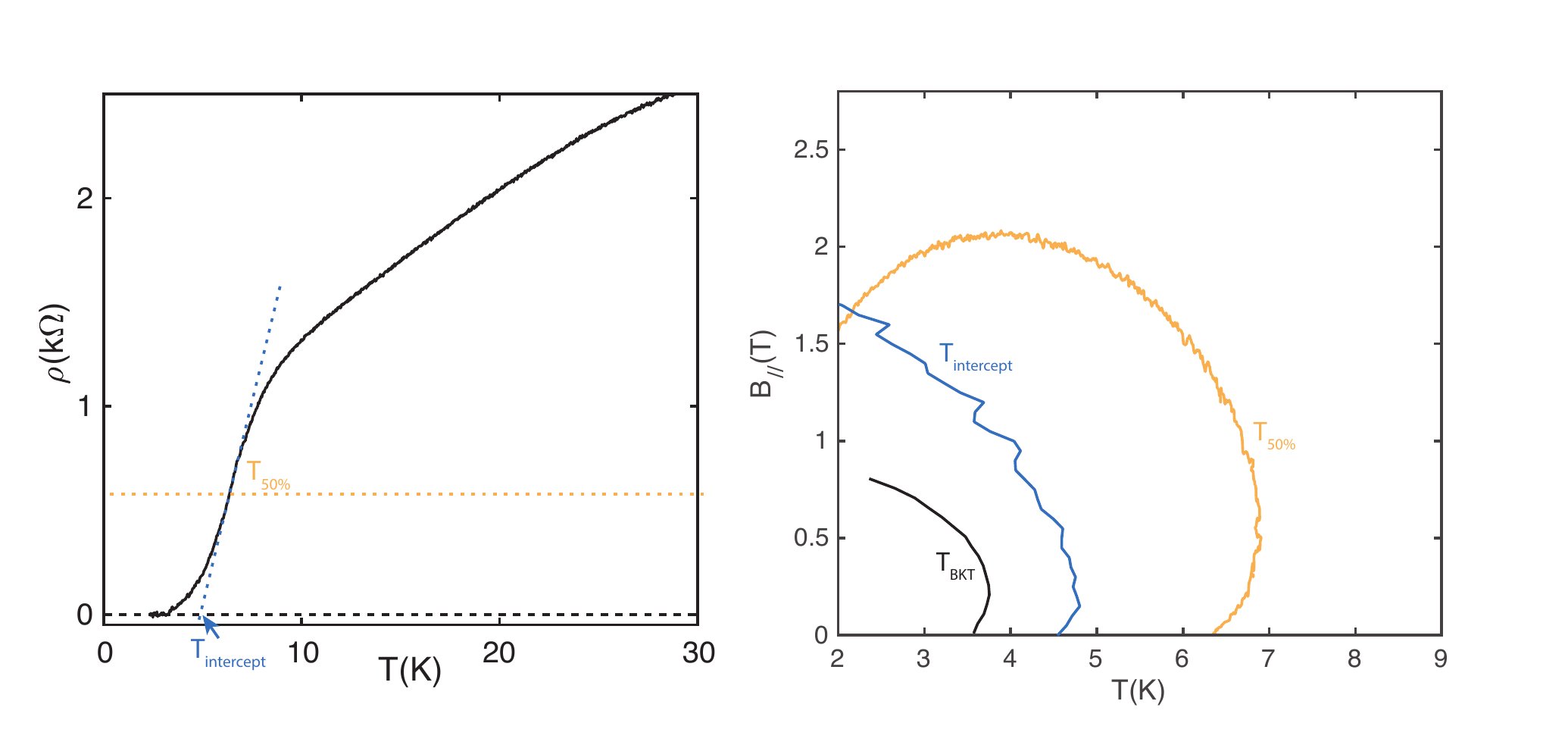}
 \caption{Critical temperature defined by different metrics. Left, a typical resistivity v.s. temperature curve. Right, different critical temperatures as a function of parallel magnetic field. }{\label{FS13}}

\end{figure}

We conducted some more analysis to demonstrate the enhancement of the superconductivity due to parallel magnetic field in the low magnetic field regime. Fig. S\ref{FS13} shows a typical resistance v.s temperature profile in the superconducting phase. We can define several transition temperatures from the $\rho$-T curve. For example, we can draw a tangent line where the slope of the $\rho$-T curve is the largest, and we can extract the $x$-intercept of the tangent line and define $T_{intercept}$, which shows a non-monotonic behavior similar to $T_{50\%}$ and $T_{BKT}$.

\begin{figure}[H]
 \centering
\includegraphics[scale=0.8]{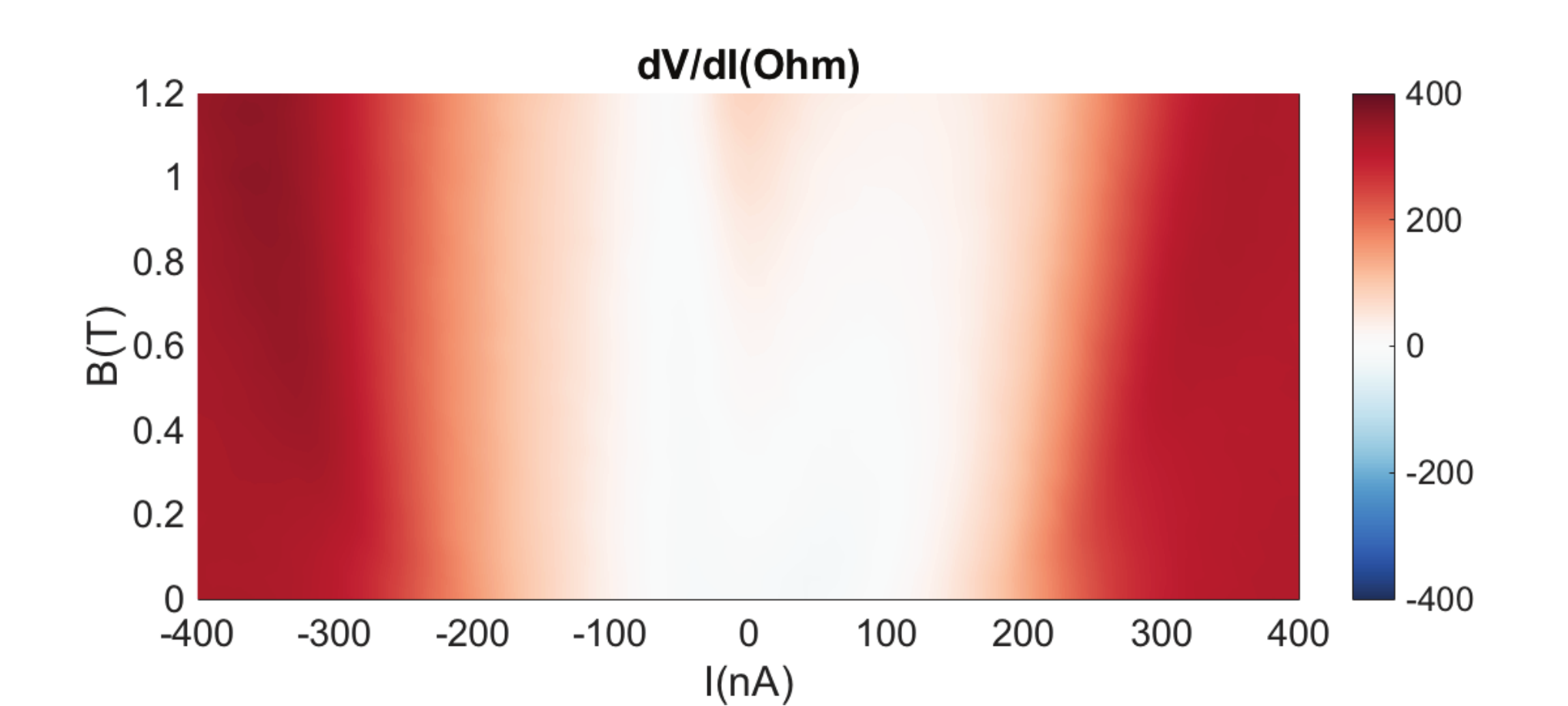}
 \caption{Differential resistance as a function of current and parallel magnetic field. The current at which resistivity starts to increase (critical current) is enhanced by a parallel magnetic field, as can be seen from the evident inverse bell shape. At even larger fields, a resistivity peak develops at zero bias current, suppressing the superconductivity.}{\label{FS14}}

\end{figure}

We can also measure differential resistance $dV/dI$ under different parallel fields. Fig. S\ref{FS14} shows clearly that the critical current (the transition between the white to red colored regimes) becomes larger as parallel field increases in the low field limit. This is another evidence that the superconductivity is strengthened by a small parallel magnetic field.

